\documentclass{article}
   \usepackage{graphicx}
   \usepackage{amssymb}
   \usepackage{epstopdf}
   \usepackage{apjgalley}
   \usepackage{multicol}
   \usepackage{psfig}
\font\fiverm=cmr5

          \font\sixrm=cmr6       
                     
\def\pr{Phys. Rev.}                             
\def\aapl{Astron. Astr. (Lett.)}                
\def\ssr{Space Sci. Rev.}                       
\def\teq#1{$\, #1\,$}                           
\def\erg{\varepsilon}

\def\dover#1#2{\hbox{${{\displaystyle#1 \vphantom{(} }\over{
   \displaystyle #2 \vphantom{(} }}$}}                

\def\sigt{\hbox{$\sigma_{\hbox{\fiverm T}}$}}                                   
\def\taupp{\tau_{\gamma\gamma}}
\def\sigpp{\sigma_{\gamma\gamma}}

\def\eb{\erg_{\hbox{\fiverm B}}}                                   

\def\ec{\erg_{\hbox{\fiverm C}}}

\def\etab{\eta_{\hbox{\fiverm B}}}
\def\etac{\eta_{\hbox{\fiverm C}}}
\def\Emax{E_{\hbox{\fiverm MAX}}}

\def\mue{\mu_{\varepsilon}}
\def\muw{\mu_{\omega}}
\def\muew{\mu_{\varepsilon\omega}}
\def\xe{x_{\varepsilon}}
\def\xw{x_{\omega}}
   
\def\chip{\chi_{{}_+}}           
\def\chim{\chi_{{}_-}}
\def\dlum{d_{\hbox{\fiverm L}}}
\def\trad{t_{\hbox{\sixrm rad}}}
\def\tdyn{t_{\hbox{\sixrm dyn}}}
\def\gammin{\gamma_{\hbox{\sixrm min}}}
\def\today{\ifcase\month\or
  January\or February\or March\or April\or May\or June\or
  July\or August\or September\or October\or November\or
  December\fi
  \space\number\day, \number\year}
\begin{document}
%
%
\newcommand{\vol}[2]{$\,$\rm #1\rm , #2.}                 
\newcommand{\figureout}[5]{\centerline{}
   \centerline{\hskip #3in \psfig{figure=#1,width=#2in}}
   \vspace{#4in} \figcaption{#5} }
\newcommand{\twofigureout}[3]{\centerline{}
   \centerline{\psfig{figure=#1,width=3.4in}
        \hskip 0.5truein \psfig{figure=#2,width=3.4in}}
    \figcaption{#3} }    
\newcommand{\figureoutpdf}[5]{\centerline{}\vskip -15pt
   \centerline{\hspace{#3in} \includegraphics[width=#2truein]{#1}}
   \vspace{#4truein} \figcaption{#5} \centerline{} }
\newcommand{\twofigureoutpdf}[3]{\centerline{}
   \centerline{\includegraphics[width=3.4truein]{#1}
        \hspace{0.5truein} \includegraphics[width=3.4truein]{#2}}
        \vspace{-0.2truein}
    \figcaption{#3} }    
\newcommand{\tableoutpdf}[5]{\centerline{}
  \vspace{#3truein}
   \centerline{\hspace{#4in} \includegraphics[width=#2truein]{#1}}
   \vspace{#5truein} \centerline{} }
\newcommand{\figureoutsubmit}[5]{
   \begin{figure*}
   \plotone{#1}
   \vspace{#4truein}
   \caption{#5} 
   \end{figure*} }   
\newcommand{\twofigureoutsubmit}[3]{
   \begin{figure*}
   \plottwo{#1}{#2}
   \caption{#3} 
   \end{figure*} }    
   
%
%
%
%

\title{TEMPORAL EVOLUTION OF PAIR ATTENUATION\\ 
        SIGNATURES IN GAMMA-RAY BURST SPECTRA}

   \author{Matthew G. Baring}
   \affil{Department of Physics and Astronomy MS-108, \\
      Rice University, P.O. Box 1892, Houston, TX 77251, U.S.A.\\
      \it baring@rice.edu\rm}
\slugcomment{To appear in \it The Astrophysical Journal, 
 Vol 649, October 1, 2006 issue.}
%

\begin{abstract}  
The spectra obtained above 100 MeV by the EGRET experiment aboard the
Compton Gamma-Ray Observatory for a handful of gamma-ray bursts has
given no indication of any spectral attenuation that might preclude
detection of bursts at higher energies.  With the discovery of optical
afterglows and counterparts to bursts in the last few years, enabling
the determination of significant redshifts for these sources, it is
anticipated that profound spectral attenuation will arise in the
Gamma-Ray Large Area Space Telescope (GLAST) energy band of 30 MeV--300
GeV for many if not most bursts. This paper explores time-dependent
expectations for burst spectral properties in the EGRET/GLAST band,
focusing on how attenuation of photons by pair creation internal to the
source generates distinctive spectral signatures.  The energy of
spectral breaks and the associated spectral indices provide valuable
information that constrains the bulk Lorentz factor of the GRB outflow
at a given time.  Moreover, the distinct temporal behavior that is
present for internal attenuation is easily distinguished from extrinsic
absorption due to intervening cosmic background fields.  These
characteristics define palpable observational goals for both
spaced-based hard gamma-ray experiments such as GLAST, and ground-based
\v{C}erenkov telescopes, and strongly impact the observability of bursts
above 300 MeV.
\end{abstract}  
\section{INTRODUCTION}
\label{sec:intro}

High energy gamma-rays have been observed for six gamma-ray bursts by
the EGRET experiment on the Compton Gamma-Ray Observatory (CGRO). Most
conspicuous among these observations is the emission of an 18 GeV photon
by the GRB940217 burst (Hurley, et al. 1994).  An additional case of
interest is provided by the so-called MILAGRITO burst (Atkins et al.
2000; Atkins et al. 2003), GRB 970417, with its uncorroborated detection
of \teq{\sim 3\sigma} significance in the TeV band. Taking into account
EGRET's field of view, its detections indicate that emission in the 1
MeV--10 GeV range is probably common among bursts, if not universal (see
Dingus 1995 for a discussion of EGRET high energy burst statistics). 
One implication of GRB observability at energies around or above 1 MeV
is that, at these energies, spectral attenuation by two-photon pair
production (\teq{\gamma\gamma\to e^+e^-}) is absent in the source.  From
this fact, Schmidt (1978) deduced that a typical burst had to be closer
than a few kpc, if it produced quasi-isotropic radiation.

In the aftermath of BATSE's revelation (e.g. Meegan et al. 1996) that
most if not all long bursts are at cosmological distances, Krolik \&
Pier (1991) and Fenimore et al. (1992) proposed that GRB photon angular
distributions are highly beamed, being produced by a
relativistically-moving plasma, a suggestion that has become an
underpinning of the GRB paradigm.  This can dramatically reduce
\teq{\taupp} below the \teq{\sim 10^{11}-10^{12}} values realized for
isotropic photons, and blueshift spectral attenuation turnovers above
the observed spectral range.  Determinations of the bulk Lorentz factor
\teq{\Gamma} of the GRB medium have mostly concentrated (e.g. Epstein
1985; Krolik \& Pier 1991; Baring 1993) on cases where the angular
extent of the source was of the order of \teq{1/\Gamma}.  These
calculations generally assume an infinite power-law burst spectrum, and
deduce (e.g. Woods \& Loeb 1995; Baring \& Harding 1997b) that gamma-ray
transparency up to the maximum energy detected by EGRET requires
\teq{\Gamma\gtrsim 100}--\teq{10^3} for cosmological bursts.  Since
BeppoSax spawned the age of precise determination of GRB redshifts, such
bounds can now be refined for the BeppoSax, HETE and Swift databases;
the discovery of high redshift bursts (e.g. see Cummings et al. 2005;
Kawai et al. 2005; Berger et al. 2006; Haislip et al. 2006, for Swift
bursts) might suggest even higher estimates of bulk Lorentz factors in
those sources.

While the power-law source spectrum assumption is expedient, the
spectral curvature seen in virtually all GRBs by BATSE (Band et al.
1993; Preece et al. 2000) is expected to play an important role in
reducing the opacity for super-GeV and TeV band emission from these
sources (Baring \& Harding 1997a). Such curvature is patently evident in
200 keV--2 MeV spectra of some EGRET-detected bursts (e.g. Schaefer, et
al. 1992), and its prevalence is indicated by the generally steep EGRET
spectra for bursts (e.g. Hurley, et al. 1994; Schneid, et al. 1992;
Sommer et al. 1994). This curvature profoundly impacts
\teq{\gamma\gamma} opacity determinations, yielding a huge dearth of
target soft photons in comparison with power-law spectra extrapolated
down to the classic X-ray band. Hence, when considering emergent photons
of dimensionless energies \teq{\erg_{\gamma}\gtrsim \Gamma^2/\eb}, where
\teq{\eb} is the dimensionless energy (in units of \teq{m_ec^2}) of the
spectral break in the BATSE band, the target photon energy is below
\teq{\eb m_ec^2} so that accurate pair opacity calculations mandate
detailed treatment of the spectral curvature observed in bursts.

This paper enunciates the principal properties of pair production
opacity that couple to spectral shape in the BATSE/EGRET energy range,
embellishing upon and focusing the work of Baring \& Harding (1997a), so
as to identify possible observational diagnostics for the Gamma-Ray
Large Area Space Telescope (GLAST; {\tt http://glast.gsfc.nasa.gov/})
mission and ground-based experiments such as the MILAGRO, MAGIC, HESS
and VERITAS atmospheric \v{C}erenkov telescopes. Moreover, it helps
define GRB science drivers for future initiatives in the hard gamma-ray
band above 100 MeV. These spectral signatures at energies where
significant opacity is realized are clearly distinguishable from
absorption by background radiation fields, and in particular via their
time-dependent evolution. The fact that turnovers produced by constant
Lorentz factor evolution can be easily discriminated from those
generated by decelerating expansions is compelling.  It may also be
possible to elicit details of the adiabaticity or otherwise of the
post-fireball expansion. Accordingly, hard \teq{\gamma}-ray telescopes
can, in principal, provide powerful probes on such evolution of bulk
motions in bright bursts, which can then be used to generate useful
bounds on the explosion energy, a key piece of information elucidating
the nature of the central engine for bursts.  Before discussing these
evolutionary effects and the potential for observational diagnostics in
Sections~\ref{sec:evolution_effect} and~\ref{sec:diagnostics}, the paper
presents refinements in the formalism for pair production opacities in
Section~\ref{sec:pairprod}.

\section{GAMMA-GAMMA PAIR PRODUCTION OPACITIES}
\label{sec:pairprod}

In this Section, a detailed formalism for \teq{\gamma\gamma} pair 
creation opacities from piecewise-continuous power-law distributions
of photons is expounded upon, extending the presentation of
Baring \& Harding (1997a) by providing greater detail and also
compact analytic reduction of the integrals for the optical depths.

\subsection{General Formalism for Internal Target Photons}
\label{sec:gen_form}

To assess the role of two-photon pair production in attenuating
gamma-ray burst spectra in the EGRET/GLAST band, the interactions of
photons created only within the emission region are considered here,
neglecting the presence of any external radiation; the impact of such
external radiation fields on burst spectra is discussed in
Section~\ref{sec:background} below.  The calculations of this
subsection mirror those of Baring (1994), whose work is used as a basis
for elucidating the effects of GRB spectral structure on pair production
opacities.  For simplicity, the photon distribution is assumed to be
azimuthally symmetric about some beaming axis (that is presumably
aligned near to the line of sight to a distant observer), with the
spectrum \teq{n(\erg )} being identical for all photon angles
\teq{\theta_{\erg}} relative to the axis within some cone of emission.
Here \teq{\erg} is the photon energy in units of \teq{m_ec^2}, a
convention adopted throughout this paper.  The photon distribution
function then takes the form \teq{n_{\gamma}(\erg ,\mue )=n(\erg )\,
f(\mue )}, with \teq{f(\mue )} normalized to unity, and where \teq{\mue
=\cos\theta_{\erg}} is the cosine of the photon angle with respect to
the axis of symmetry.  Further, for anisotropic (beamed) photons we
take the simple case that the angular distribution about the axis is
uniform within a cone of half-angle \teq{\theta_{\erg}=\theta_m}:
\begin{equation}
   f(\mue )\; =\;\dover{1}{1-\mu_m}\quad ,\quad \mu_m\,\equiv\,\cos\theta_m
   \,\leq\,\mue\,\leq\, 1\quad .
 \label{eq:angle_dist}
\end{equation}
This conical sector form is not exactly equivalent to the angular
distribution of an isotropic photon distribution boosted by some bulk
Lorentz factor \teq{\Gamma\sim 1/\theta_m} with respect to an observer
at infinity, since it lacks the ``wings'' at \teq{\mue\lesssim \mu_m}.
Yet it suffices to exhibit the general nature of photon beaming effects
on pair production rates.  Consequently, the results presented here, for
example that inferred from Eq.~(\ref{eq:calT_beam}), differ slightly
from what would be derived from a beamed adaptation of the isotropic
results of Gould and Schreder (1967, as expounded briefly below), which
was the approach of Krolik \& Pier (1991) and Fenimore, Epstein \& Ho
(1992). Scientific conclusions pertaining to relativistic beaming in
GRBs are not contingent upon such subtleties, largely because the
computed optical depths are most sensitive to the beaming factor
\teq{\theta_m}, i.e. \teq{1/\Gamma}, among the relevant parameters.
Moreover, adherence to one particular angular distribution is not
mandated by observations nor by theoretical insights. For example,
diffusive particle acceleration at the relativistic shocks that are
suggested as being sites for energization in gamma-ray bursts does not
generate isotropic distributions at the shock in any frame of reference
(e.g. see Bednarz \& Ostrowski 1998; Kirk et al. 2000).

For photons that interact with themselves to produce pairs, a slight
modification of the form for the pair production optical depth
\teq{\taupp (\erg )} in Eq.~(7) of Stepney and Guilbert (1983) was
adopted by Baring (1994) as a starting point for analytic developments.
This form, appropriate for photon distributions independent of azimuthal
angles, is
\begin{eqnarray}
   \taupp (\erg ) &=& \dover{4R}{\pi}\int_{-1}^1 d\mue\, f(\mue )
   \int_{-1}^1 d\muw\, f(\muw ) \vphantom{\Biggl(}
   \int_0^{\infty} d\omega\, \dover{n(\omega )}{\erg\omega}\nonumber\\[-5.5pt]
 \label{eq:taupp} \\[-5.5pt]
   & &\quad\vphantom{\Biggl(}\times\int_{\chim}^{\chip} 
   \dover{\chi^3\,\sigpp (\chi )\, d\chi}{\sqrt{\bigl( \chip^2-\chi^2\bigr)\,
   \bigl(\chi^2-\chim^2\bigr)}} \quad .\nonumber
\end{eqnarray}
Here \teq{\chi =\lbrack\erg\omega (1-\cos\Theta)/2\rbrack^{1/2}} is the
center-of-momentum (CM) frame energy of the photons, whose energies 
and angle cosines in the observer's frame are \teq{(\erg ,\,\mue )} and
\teq{(\omega ,\,\muw)} respectively.  \teq{\Theta} is the angle between
the photon directions.  The pair production threshold condition is then
\teq{\chi\geq 1}.  Since \teq{\mue =\cos\theta_{\erg}} and \teq{\muw
=\cos\theta_{\omega}}, the range \teq{[0,\, \pi]} of azimuthal
angles in the observer's frame establishes the bounds
\begin{equation}
   \chi^2_{{}_{\pm}}\; =\;\dover{\erg\omega}{2}\,\Bigl\{ 1-
   \cos (\theta_{\erg}\pm\theta_{\omega})\Bigr\}\quad . 
 \label{eq:chisq}
\end{equation}

The focus of Baring (1994) was on infinite power-law {\it number
density} distributions \teq{n(\omega )}.  In this paper, low energy
cutoffs with \teq{\omega >0} will be introduced, since these prove
expedient in treating piecewise continuous broken power-laws, which
constitute suitable approximations to the general shape of burst
spectra.  Specifically, the form
\begin{equation}
   n(\erg )\; =\; \cases{ 0\;\; , & \teq{\quad \erg < \ec\vphantom{\bigl(}}\cr
                        n_{\gamma} \,\erg^{-\alpha}
                        \;\; , & \teq{\quad \erg > \ec\vphantom{\bigl(}}\cr}
 \label{eq:ndef}
\end{equation}
is used for some minimum photon energy \teq{\ec m_ec^2}, following Gould
and Schreder (1967) and Baring \& Harding (1997a); note that the works
of Baring (1993, 1994) used an integral photon flux index. The units of
both \teq{n_{\gamma}} and \teq{n(\erg )} correspond to inverse volumes. 
The change of variables \teq{\zeta = (1-\cos\Theta )/2
=\chi^2/\erg\omega} can then be employed to facilitate the manipulation
of the integrals along the lines of Baring (1994).  Defining a parameter
\begin{equation}
   \eta\; =\; \sqrt{\erg\ec}\quad ,
 \label{eq:etadef}
\end{equation}
that is the center-of-momentum frame energy for head-on photon collisions,
the form
\begin{equation}
   \taupp (\erg )\; =\; n_{\gamma}\sigt R\,
   {\cal T}_{\alpha}\Bigl(\theta_m \, ,\;
        \sqrt{\erg\ec}\,\Bigr)\,\erg^{\alpha -1}\quad ,
 \label{eq:tauppform}
\end{equation}
is obtained, where 
\begin{equation}
   {\cal T}_{\alpha}\Bigl(\theta_m \, ,\; \eta \,\Bigr) \; \equiv\;
   \dover{4}{\sigt}\int_1^{\infty} d\chi\,
   \dover{\sigpp (\chi )}{\chi^{2\alpha -1}}\;
   {\cal F}_{\alpha}(\theta_m,\, \eta, \, \chi)
 \label{eq:calTdef}
\end{equation}
is an integration over the pair production cross-section \teq{\sigpp}, and 
the angular distributions contribute the factor
\begin{eqnarray}
   {\cal F}_{\alpha}(\theta_m,\, \eta, \, \chi) & = &\dover{1}{\pi}
   \int_{\mu_m}^1 \dover{d\mue}{1-\mu_m}\int_{\mu_m}^1 \dover{d\muw}{1-\mu_m}
   \nonumber\\[-5.5pt]
 \label{eq:calFdef} \\[-5.5pt]
   && \int_{\zeta_-}^{\zeta_+} \dover{\zeta^{\alpha}\,d\zeta}{
   \sqrt{(\zeta_+-\zeta )(\zeta -\zeta_-)}}\;
   \Theta \biggl( \dover{\eta^2\zeta}{\chi^2} \biggr)
   \;\; .\nonumber
\end{eqnarray}
Here, \teq{\Theta (x)} is a Heaviside step function such that
\teq{\Theta (x)=1} for \teq{0\leq x\leq 1} and is zero otherwise.  It
expresses the kinematic condition \teq{\chi^2\geq \erg\ec\zeta} for
\teq{\zeta = (1-\cos\Theta )/2} that is imposed by the spectral
truncation.  The \teq{\zeta}-integration limits are
\teq{\zeta_{\pm}=\chi^2_{{}_{\pm}}/\erg\omega =\lbrack 1-\cos
(\theta_{\erg}\pm\theta_{\omega})\rbrack /2}. Note that in the case of
\teq{\ec =0}, effecting the substitution \teq{\alpha \to \alpha +1} in
these results reproduces those of Baring (1994).

It is elucidating to focus first on the case of isotropic photons, i.e.
when \teq{\theta_m=\pi}.  The angular integrations  in
Equation~(\ref{eq:calFdef}) are facilitated by following Baring (1994)
and changing variables to \teq{\phi} with \teq{2\zeta =
\{\zeta_++\zeta_-+(\zeta_+-\zeta_-)\cos\phi\}}, and then performing the
solid angle transformation \teq{(\muw,\,\phi )\to (\muew
,\,\phi_{\erg\omega})}, where \teq{\muew=\cos\Theta} is the cosine of
the angle between the photons.  This results in a dramatic
simplification of the three integrals.  Defining
\begin{equation}
   q \; =\; \min \biggl\{ 1\, ,\; \dover{\chi}{\eta}\, \biggr\}\quad ,
 \label{eq:qdef}
\end{equation}
Equation~(\ref{eq:calFdef}) becomes, in the limit \teq{\theta_m\to\pi},
\begin{equation}
   {\cal F}_{\alpha}(\pi,\, \eta, \, \chi)\; =\; \dover{q^{2(\alpha +1)}}{
   \alpha +1}\quad ,
 \label{eq:calFisotropy}
\end{equation} 
with the \teq{\eta \leq 1} case (i.e. infinite power-law; \teq{q=1}) 
reproducing Eq.~(10) of Baring (1994).  It then follows that
\begin{equation}
   {\cal T}_{\alpha}\bigl(\pi \, ,\; \eta \,\bigr) \; =\;
   \dover{{\cal H}(\alpha ,\, \eta )}{\alpha + 1}
 \label{eq:calTisotropy}
\end{equation}
where 
\begin{equation}
   {\cal H}(\alpha ,\,\eta )\; =\;\dover{4}{\sigt}\int_1^{\infty}
   \dover{q^{2(\alpha +1)}}{\chi^{2\alpha -1}} \;\sigpp (\chi )\, d\chi\quad .
 \label{eq:calHdef}
\end{equation}
This completely defines the pair production optical depth for the case
of an isotropic distribution of photons with a power-law spectrum that is
truncated at low energies, a result that is quickly shown to be
equivalent to Eqs.~(21)--(23) of Gould and Schreder (1967) by observing
their notation \teq{s\to \chi^2}, and by performing the appropriate
integration by parts.  

An exact expression for \teq{{\cal T}_{\alpha} (\pi ,\, 1)} was obtained by 
Svensson (1987; see his Equation [B6]), and can be approximated (Baring
1993, 1994) to better than 1\% for \teq{1 < \alpha < 7} using \teq{{\cal
H}(\alpha,\, 1)\approx 7/6/\alpha^{5/3}}.  An asymptotic form for
\teq{\eta\to\infty} can be obtained most expediently by adopting a form
equivalent to Equation~(\ref{eq:calTisotropy}):
\begin{equation}
   {\cal T}_{\alpha}\bigl(\pi\, ,\; \eta \,\bigr) \; \equiv\;
   \dover{8}{\sigt}\int_{\eta}^{\infty} \dover{d\chi}{\chi^{2\alpha +3}}\,
   \int_1^{\chi} x^3\, \sigpp (x)\, dx\quad .
 \label{eq:calTaltern}
\end{equation}
Using the asymptotic form \teq{\sigpp (x) \approx (3\sigt /8)\,
[2\log_e2x-1]/x^2} for \teq{x\gg 1}, i.e. for the extreme Klein-Nishina domain, 
one quickly arrives at
\begin{equation}
   {\cal T}_{\alpha}\bigl(\pi\, ,\; \eta \,\bigr) \; \approx\;
   \dover{3}{2\alpha}\; \dover{1}{\eta^{2\alpha}}\;
   \biggl\{ \log_e2\eta -1 + \dover{1}{2\alpha} \biggr\}\;\; ,\quad \eta\;\gg\; 1\;\; .
 \label{eq:calTapprox_KN}
\end{equation}
This is identical to Eq.~(24a) of Gould and Schreder (1967).

\subsection{Angular Contributions for Strong Beaming}
\label{sec:angular_cont}

Exploration of the angular integrals for cases \teq{\mu_m\neq -1}, and 
specifically those where beaming is strong and \teq{\mu_m\approx 1}, 
is not as straightforward as the isotropic photon situation.
Equation~(\ref{eq:calFdef}) must be manipulated considerably to 
extract compact analytic forms.  In the limit \teq{1-\mu_m\approx \theta_m^2/2\ll 1}, 
to leading order in \teq{(1-\mu_m)},  it is easily shown that 
\teq{\zeta_{\pm}\approx (\,\sqrt{1-\mue} \pm \sqrt{1-\muw}\, )^2/2}, suggesting 
the change of integration variables to \teq{\kappa} such that 
\teq{\zeta =\kappa^2 (1-\mu_m)/2} together with \teq{1-\mue = (1-\mu_m) \xe^2} 
and \teq{1-\muw = (1-\mu_m) \xw^2}.  This yields the following form 
for \teq{{\cal F}_{\alpha}}:
\begin{eqnarray}
   && \hbox{\hskip -8pt}
   {\cal F}_{\alpha}(\theta_m,\, \eta, \, \chi) \; \approx \;
   \dover{16\,\theta_m^{2\alpha}}{2^{2\alpha}\pi}\; 
   \int_0^1 \xe\, d\xe  \int_0^{\xe} \xw\, d\xw \nonumber \\[-5.5pt]
 \label{eq:calFapprox}\\[-5.5pt]
   && \hbox{\hskip -8pt} \times \;\int_{\kappa_-}^{\kappa_+} 
   \dover{ \kappa^{2\alpha+1}\, d\kappa}{
      \sqrt{ [(\kappa_+)^2-\kappa^2]\; [\kappa^2 - (\kappa_-)^2]}}\;\;
   \Theta \biggl( \dover{\kappa\eta\theta_m}{2\chi} \biggr)\; , \;\;
   \theta_m\ll 1\, ,\nonumber
\end{eqnarray}
for \teq{\kappa_{\pm} = \xe\pm\xw}.  Here, the fact that the integration
is symmetric under the interchange \teq{\xe\leftrightarrow\xw} has been
used to write \teq{\xw\leq\xe} without loss of generality, thereby
introducing an extra factor of two.

Before evaluating Eq.~(\ref{eq:calFapprox}) in generality, it is
instructive to identify compact results in two asymptotic regimes. In
the particular case of infinite power-law spectra, where \teq{\eta =0},
the value of \teq{{\cal F}_{\alpha}} is independent of \teq{\chi} so
that the angular integrations separate from the energy ones.  Then,
further changes of variables along the lines of \teq{\xw = w\xe} and
\teq{\kappa^2 = \xe^2 [ (1-w)^2+4 w z ]} leads to evaluation of the
integrals using identities 3.197.3, 9.134.3 and 7.512.4 from Gradshteyn
\& Rhyzik (1980):
\begin{eqnarray}
   {\cal F}_{\alpha}(\theta_m ) &\approx & {\cal A}(\alpha )
   \,\theta_m^{2\alpha}\quad , \nonumber\\[-5.5pt]
 \label{eq:calAdef}\\[-5.5pt]
   {\cal A}(\alpha ) &=& \dover{2^{1-2\alpha}\Gamma (2\alpha +2)}{
\Gamma (\alpha +2)\,\Gamma (\alpha +3)} \quad ,\nonumber
\end{eqnarray}
valid for \teq{\eta\theta_m/\chi\ll 1}. This result is precisely that in
Eq.~(12) of Baring (1994) after the adjustment for the different
spectral index convention used there; Baring (1994) also posited an
approximation, \teq{{\cal A}(\alpha ) \approx 2/(4/3+\alpha )^{27/11}}
that is accurate to better than 1\% on \teq{1 <\alpha <7}. 
Equation~(\ref{eq:calAdef}) implies
\begin{equation}
   {\cal T}_{\alpha}(\theta_m ,\, \eta ) \;\approx\; {\cal A}(\alpha )\,
   {\cal H}(\alpha ,\, 1)\;\theta_m^{2\alpha}\;\; , \quad
   \eta\theta_m\;\leq\; 1\;\; .
 \label{eq:calT_beam}
\end{equation}
This can be inserted into Eq.~(\ref{eq:tauppform}) to obtain the 
corresponding final result for the pair production optical depth for cases 
where the beaming is strong, but when the spectral cutoff of the photon 
distribution does not come into play.

The opposite asymptotic limit arises for \teq{\eta\theta_m/\chi\gg 1},
and is readily tractable since it samples \teq{\kappa\ll 1} and
\teq{\xe\approx\xw} domains.  This results in
\begin{equation}
   {\cal F}_{\alpha}(\theta_m,\, \eta, \, \chi) \;\approx\; \dover{4}{\alpha +1}
   \,\theta_m^{2\alpha}\; \biggl\{ \dover{\chi}{\eta\theta_m}
   \biggr\}^{2(\alpha +1)} , \quad \dover{\eta\theta_m}{\chi}\gg 1.
 \label{eq:calF_largeeta}
\end{equation}
Comparison with Eq.~(\ref{eq:calAdef}) indicates that the coefficients
for the \teq{\eta\theta_m \ll\chi} and \teq{\eta\theta_m \gg\chi}
domains differ, largely due to the behavior of \teq{{\cal F}_\alpha}
near the critical kinematic value of \teq{\eta\theta_m =\chi}. Observe
that this asymptotic domain establishes \teq{{\cal
F}_{\alpha}(\theta_m,\, \eta, \, \chi) \propto 1/\theta_m^2}, which is
essentially a phase space factor in the angular integrals.

To reduce Eq.~(\ref{eq:calFapprox}) for general values of
\teq{\eta\theta_m/\chi}, a moderate amount of manipulation is required. 
An effective approach is to change the integration variables via
\teq{x_{\omega} = w x_{\erg}} and \teq{\kappa = x_{\erg}\rho}, so that
\teq{x_{\erg}}, \teq{w} and \teq{\rho} constitute the new set of
variables. The \teq{w} integration is then performed first, leading to
the appearance if a simple inverse trigonometric function.  The
procedure is detailed in the Appendix, where it is found that \teq{{\cal
F}_{\alpha}} is expressible in terms of elementary functions and the
hypergeometric function \teq{F\equiv {}_2F_1}.  The attractively compact
result of these manipulations is (for \teq{\theta_m\ll 1})
\begin{equation}
   {\cal F}_{\alpha}(\theta_m,\, \eta, \, \chi) \; \approx\; \theta_m^{2\alpha}\; 
        \cases{ {\cal A}(\alpha)\; , & $\quad\Psi\;\geq\; 1 \;\; \vphantom{\Bigl(}$; \cr
                      {\cal G}_{\alpha}(\Psi )\; , & $\quad 0\;\leq\; \Psi\;\leq\; 1\;\; 
                                 \vphantom{\Bigl(}$  , \cr}
 \label{eq:calF_final}
\end{equation}
where \teq{\Psi =\chi/(\eta\theta_m)}, and \teq{{\cal G}_{\alpha}(\Psi )} 
is given by Eq.~(\ref{eq:calG_final}):
\begin{eqnarray}
     {\cal G}_{\alpha}(\Psi) &=& \dover{8}{\pi}\, \Psi^{2(1+\alpha )}
           \Biggl\{ \dover{\arccos\Psi}{\alpha +1}
                 - \dover{\Psi\sqrt{1-\Psi^2}}{\alpha +2}\nonumber\\[-5.5pt]
  \label{eq:calG_enunc}\\[-5.5pt]
                 &&\qquad +\dover{\Psi\, F(1/2,\, \alpha+ 3/2;\, \alpha + 5/2;\, \Psi^2)}{
                                (\alpha +1)\, (2\alpha +3)\, (\alpha +2)} \Biggr\}\quad .\nonumber
\end{eqnarray}
Accordingly, much of the character of \teq{{\cal F}_{\alpha}}
is encapsulated in the one parameter \teq{\Psi}.  Numerical protocols
for the evaluation of the hypergeometric function
\teq{F(1/2,\, \alpha+ 3/2;\, \alpha + 5/2;\, \Psi^2)} are outlined in the
Appendix.  Observe that \teq{{\cal G}_{\alpha}(\Psi)\leq {\cal A}(\alpha)}
for \teq{0\leq\Psi\leq 1}.  Using this formalism, Eq.~(\ref{eq:calTdef}) 
can now be written
\begin{equation}
   {\cal T}_{\alpha}(\theta_m,\, \eta ) \; \approx\; \theta_m^{2\alpha}\; 
   {\cal K} (\alpha ,\, \eta\theta_m )\quad ,
 \label{eq:calT_calKform}
\end{equation}
where
\begin{eqnarray}
   {\cal K}(\alpha ,\,\eta ) & = & \dover{4}{\sigt}\int_1^{\hat{\eta}}
       d\chi\, \dover{\sigpp (\chi )}{\chi^{2\alpha -1}} 
       \; {\cal G}_{\alpha}\Bigl(\dover{\chi}{\hat{\eta}}\Bigr)\nonumber\\[-5.5pt]
 \label{eq:calKdef}\\[-5.5pt]
   & + & {\cal A}(\alpha ) \int_{\hat{\eta}}^{\infty}
       d\chi\, \dover{\sigpp (\chi )}{\chi^{2\alpha -1}} \;\; , \quad
       \hat{\eta}\; =\; \max \Bigl\{ 1,\; \eta \Bigr\} \;\; .\nonumber
\end{eqnarray}
The general character of \teq{{\cal K}(\alpha ,\,\eta )} is illustrated
in Fig.~\ref{fig:calK_char}, including its comparison with the similar
function \teq{{\cal H}(\alpha ,\,\eta )} that pertains to the isotropic
photon case above.

Further attempts to reduce the integration over \teq{\chi} in the
expression for \teq{{\cal T}_{\alpha}} lead to undesirable algebraic
complexity, introducing somewhat obscure special functions that preclude
further insight.  Hence numerical evaluations of the integral over
\teq{\chi} suffice.  These can be checked using two identifiable
asymptotic formulae.  The first is the \teq{\eta\theta_m\leq 1} form
that is encapsulated in Eq.~(\ref{eq:calT_beam}), which uses \teq{{\cal
K}(\alpha ,\, \eta )\to {\cal K}(\alpha ,\, 1) \equiv {\cal A}(\alpha
)\, {\cal H}(\alpha ,\, 1)} for \teq{\eta\leq 1}.  The second is for the
extreme Klein-Nishina limiting case of \teq{\eta\theta_m\gg 1}, which
can be obtained by inserting Eq.~(\ref{eq:calF_final}) into
Eq.~(\ref{eq:calTdef}), and then invoking the integral evaluations in
Eq.~(\ref{eq:int_calG_idents}). The result is
\begin{equation}
   {\cal T}_{\alpha}(\theta_m ,\, \eta ) \;\approx\; 
        \dover{3}{2\alpha}\,\dover{1}{\eta^{2\alpha}}\;
        \Bigl\{ \log_e\eta\theta_m   - \dover{3}{4}+ \dover{1}{2\alpha} \Bigr\}
        \; ,\quad \eta\theta_m\;\gg\; 1\; ,
 \label{eq:calTbeam_asymp}
\end{equation}
which provides an excellent approximation to \teq{{\cal
T}_{\alpha}(\theta_m ,\, \eta )} for \teq{\eta\theta_m\gtrsim 4}. The
similarity of this to the Klein-Nishina asymptotic form for isotropic
photons in Eq.~(\ref{eq:calTapprox_KN}) is noticeable, highlighting how
in this domain, the extreme kinematic collimation imposed by the
\teq{\chi^2\geq \eta^2 (1-\cos\Theta )/2} condition is more constraining
than the beaming of the photon distribution within an angle
\teq{\theta_m}.  This characteristic is evinced in
Fig.~\ref{fig:calK_char} via the approximate constancy of the ratio
\teq{{\cal K}(\alpha ,\,\eta )/{\cal H}(\alpha ,\,\eta )} for large
\teq{\eta}: comparison of Eq.~(\ref{eq:calTbeam_asymp}) with
Eq.~(\ref{eq:calTapprox_KN}) reveals that \teq{{\cal K}(\alpha ,\,\eta
)/{\cal H}(\alpha ,\,\eta )\to {\cal A}(\alpha )/(\alpha +1)} as
\teq{\eta\to\infty}.

\figureoutpdf{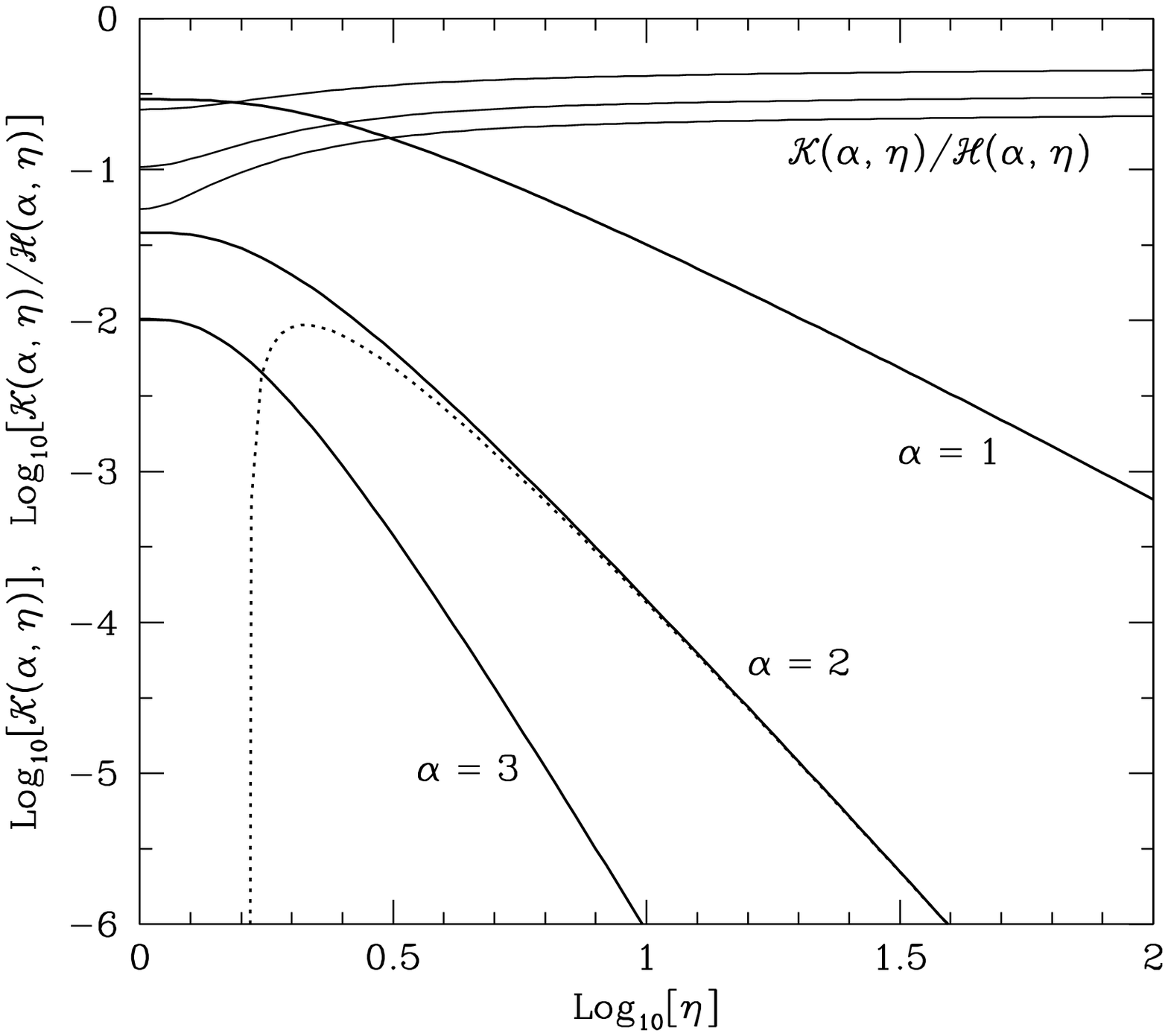}{3.7}{0.0}{-0.2}{
The general behavior of \teq{{\cal K}(\alpha ,\,\eta )} is depicted as a function
of \teq{\eta} as the heavy-weight curves, for three values of the spectral index
\teq{\alpha}, as indicated.  Also exhibited as the dotted curve is the asymptotic 
approximation for \teq{{\cal K}(\alpha ,\,\eta )} for \teq{\eta\gg 1}, precisely
Eq.~(\ref{eq:calTbeam_asymp}) with the substitution \teq{\theta_m\to 1}.
To compare with the function \teq{{\cal H}(\alpha ,\,\eta )} that appears in the
formalism for isotropic photons, in the upper portion of the plot
the values of the ratio \teq{{\cal K}(\alpha ,\,\eta )/{\cal H}(\alpha ,\,\eta )}
are illustrated as lighter weight curves, specifically for the indices
\teq{\alpha =1,2,3} ranging from top to bottom.
 \label{fig:calK_char}} 

For the practitioner desiring an even more compact approximation for the
optical depth, a useful result that embodies the essence of both the two
beamed forms in Eqs.~(\ref{eq:calT_beam}) and~(\ref{eq:calTbeam_asymp})
can be put forward. There is no unique choice, however a comparison of
the coefficients in Eqs.~(\ref{eq:calAdef}) and~(\ref{eq:calF_largeeta})
suggests the following mapping that mimics the appearance of the
isotropic form:
\begin{eqnarray}
   {\cal F}_{\alpha}(\theta_m,\, \eta, \, \chi) &\to & {\cal A}(\alpha )
   \, Q^{2(\alpha+1)}\,\theta_m^{2\alpha}\quad , \nonumber\\[-5.5pt]
 \label{eq:calFhybrid}\\[-5.5pt]
   Q &=& \min \biggl\{ 1\, ,\; \dover{\chi}{{\cal B}(\alpha )\,\eta\theta_m}\, \biggr\}
   \quad .\nonumber
\end{eqnarray}
Insertion of this into Eq.~(\ref{eq:calTdef}) then yields a development
identical to that leading to Eqs.~(\ref{eq:calTisotropy}) and~(\ref{eq:calHdef}),
but with the substitution \teq{\eta\to {\cal B}(\alpha )\,\eta\theta_m}.
Hence, in cases where \teq{{\cal B}(\alpha )\,\eta\theta_m \leq 1},
this protocol automatically generates the result in Eq.~(\ref{eq:calT_beam}).
Comparison of Eq.~(\ref{eq:calTapprox_KN}), with 
\teq{\eta\to {\cal B}(\alpha )\,\eta\theta_m}, and Eq.~(\ref{eq:calTbeam_asymp}) 
then establishes the correspondence
\begin{equation}
   {\cal B}(\alpha )\; =\; \Bigl[ (\alpha + 1)\, {\cal A}(\alpha ) \Bigr]^{1/(2\alpha )}
 \label{eq:calBdef}
\end{equation}
from these Klein-Nishina limiting forms, and the resulting compact 
approximation
\begin{equation}
   {\cal T}_{\alpha}(\theta_m ,\, \eta ) \;\approx\; {\cal A}(\alpha )\,
   {\cal H}\Bigl(\alpha ,\, {\cal B}(\alpha )\,\eta\theta_m\Bigr)\;\theta_m^{2\alpha}\;\; 
 \label{eq:calT_alt}
\end{equation}
is generally accurate to around 10--20\% for \teq{\eta\theta_m\gtrsim
1}. Note that \teq{{\cal B}(\alpha )} is of the order of magnitude of
unity for typical \teq{\alpha}. This approximation essentially maps over
to a correspondence of \teq{{\cal B}(\alpha )\theta_m\leftrightarrow
1/\Gamma} between beaming angle and bulk Lorentz factor \teq{\Gamma} of
the emission region.  Clearly such an approximation is subjective, but
does not compromise the appropriateness of pair production attenuation
calculations.  Notwithstanding, for the investigation here,
Eqs.~(\ref{eq:calT_calKform}) and~(\ref{eq:calKdef}) will form the basis
of the calculations below.

\subsection{General Constructs for Complicated GRB Spectra}
 \label{sec:gen_GRB_spectra}
 
The adopted form for the approximate optical depth due to pair creation
from a beamed conical power-law photon distribution with a low energy 
cutoff at dimensionless energy \teq{\ec} (generally \teq{\ll 1}) is
\begin{equation}
   \taupp (\erg )\; =\; n_{\gamma}\sigt R\, 
   {\cal K}\Bigl( \alpha ,\, \theta_m\,\sqrt{\erg\ec}\Bigr)
   \;\theta_m^{2\alpha}\;\erg^{\alpha -1}\quad .
 \label{eq:taupp_beam_approx}
\end{equation}
Here, given that the angular form in Eq.~(\ref{eq:angle_dist}) is
normalized to unity, the normalization parameter \teq{n_{\gamma}},
defined in Eq.~(\ref{eq:ndef}), is such that
\teq{n_{\gamma}\theta_m^{2\alpha}/(\alpha-1)} represents the total
photon number density above the pair production threshold \teq{\erg\sim
1/\theta_m} for the beamed photon distribution. The expression
corresponding to Eq.~(\ref{eq:taupp_beam_approx}) for a power-law that
extends down to zero energy but possesses a high-energy cutoff is
obviously obtained by subtracting this result from that for an infinite
power-law, namely Eq.~(\ref{eq:calT_beam}).  It is then a simple matter
to construct optical depths for more complicated burst spectral
structure.  The particular spectrum that will be adopted throughout this
paper is that used in Baring \& Harding (1997a; hereafter BH97a), namely
a power-law broken at a dimensionless energy \teq{\eb} with a low energy
cut-off at \teq{\ec}:
\begin{equation}
   n(\erg )\; =\; n_{\gamma}\eb^{-\alpha_h}\, \cases{
   0, & if \teq{\erg\leq\ec\;\;},\cr
   \eb^{\alpha_l}\erg^{-\alpha_l} , &
     if \teq{\vphantom{\Bigl(} \ec\leq\erg\leq\eb\;\;},\cr
   \eb^{\alpha_h}\erg^{-\alpha_h} , & \teq{\erg >\eb\;\; }. \cr }
 \label{eq:powerlaw}
\end{equation}
This form mimics the generic shape of burst spectra (see Table~1 below
for a listing of pertinent spectral parameters for BATSE/EGRET bursts),
and the introduction of a sharp cutoff at \teq{\ec} is to encompass
scenarios where there is severe soft photon paucity.  Note that typical
burst radiation mechanisms such as synchrotron emission and inverse
Compton scattering from quasi-isotropic particles formally generate
spectra with \teq{\ec\to 0}, though in practice, the inverse Compton
mechanism, so-called pitch angle synchrotron emission (e.g. see Epstein
1973), and moreover self-absorbed situations (e.g. Crider \& Liang 1999;
Granot, Piran \& Sari 2000), produce extremely flat spectra so that
finite choices of \teq{\ec} can be realistic.  Using the formalism
described above, the optical depth for pair production attenuation of
such a broken power-law/truncated photon distribution is found to be
\begin{eqnarray}
   \dover{\taupp (\erg )}{n_{\gamma}\sigt R} &\approx &
   \theta_m^{2\alpha_l}\; \Bigl\{ {\cal K}(\alpha_l,\, \etac) - {\cal K}(\alpha_l,\, \etab )\Bigl\}\,
   \dover{\erg^{\alpha_l-1}}{\eb^{\alpha_h-\alpha_l}}
     \nonumber\\[-5.5pt]
 \label{eq:taupp_final}\\[-5.5pt]
   &&  \qquad + \;\; \theta_m^{2\alpha_h}\; {\cal K}(\alpha_h,\, \etab )\,
    \erg^{\alpha_h-1} \quad ,\nonumber
\end{eqnarray}
where 
\begin{equation}
   \etab \; =\; \theta_m\sqrt{\eb\erg}  \quad , \quad
   \etac \; =\; \theta_m\sqrt{\ec\erg} \;\; .
 \label{eq:etabetac_def}
\end{equation}
Observe that at \teq{\erg\sim 1/(\eb\theta_m^2)} the two spectral terms
in Eq.~(\ref{eq:taupp_final}) become comparable, indicating that this
energy is the domain where the attenuation ``image'' of the the
BATSE-band spectral break at \teq{\eb} is experienced. This expression
for the optical depth, to be used hereafter in the computations of this
paper, closely resembles that in Eq.~(2) of BH97a, differing only by its
use of \teq{{\cal K}} functions in the place of \teq{{\cal H}} functions
and the replacement \teq{\Gamma\to 1/\theta_m}; ensuing results in this
paper are quantitatively but not qualitatively different from those in
BH97a.

Clearly, more gradual spectral curvature can be treated by fitting the
GRB continuum with piecewise continuous power-laws, generalizing the
structure inherent in Eq.~(\ref{eq:taupp_final}) to sums of various
power-law factors combined with differences in \teq{{\cal K}} functions.
 The technique presented here is well suited to this adaptation, though
for typical burst spectral forms, only modest changes in \teq{\taupp}
will be afforded by such refinement of the optical depth calculation,
and then only deep into the attenuation trough. There may exist, of
course, occasional pathological exceptions that require such refinement.

\subsection{Cosmological Background Targets}
 \label{sec:background}

The cosmic background infra-red (IR) and optical starlight fields
provide target photons external to gamma-ray bursts that can attenuate
in the EGRET/GLAST band.  They effect line-of-sight absorption, and so
are exponential functions of the line-of-sight optical depth.  Moreover,
they are independent of burst characteristics such as flux, variability
and spectral evolution, and therefore are markers only of the distance
or redshift to a given burst.  Such a property is touted as a means of
probing the background fields using any bursts (and active galaxies)
detected by the atmospheric \v{C}erenkov technique.  The attenuation
effect of intervening cosmic background fields on hard gamma-rays via
\teq{\gamma\gamma\to e^+e^-} has been extensively studied in the context
of active galaxies (e.g. Stecker, de Jager \& Salamon 1992, MacMinn \&
Primack 1996; for reviews, see Primack et al. 2001, Stecker 2001), with
predictions being sensitive to the assumed level of the infra-red
background, a historically poorly measured quantity. This uncertainty
has been pervasive in the discussion of absorption of TeV photons from
active galaxies, predominantly for AGNs that are detected at TeV
energies by atmospheric \v{C}erenkov telescopes have been nearby (e.g.
Mrk 421 and Mrk 501 are at \teq{z=0.031} and \teq{z=0.034},
respectively). However, this discussion has experienced a profound
development in the recent observation (Aharonian et al. 2006) by the
HESS telescope array of harder than expected TeV-band spectra in two
more distant blazars, H 2356-309 at \teq{z=0.165} and 1ES 1101-232 at
\teq{z=0.186} (see also Albert et al. 2006b, for the very recent MAGIC
detection of 1ES 1218+30.4 at \teq{z=0.182}). This extension to more
distant sources has reduced the upper bounds on the extragalactic
background IR light to interesting levels that are within 50\% of lower
limits set by galaxy counts with Hubble Space Telescope data.  This
advance moves the field of TeV gamma-ray astronomy much closer to being
a powerful probe of these radiation backgrounds.

Line-of-sight attenuation of GRB spectra provides a very different case,
at least for those bursts at moderate to high redshifts that were seen
by the BeppoSax mission, and are currently being detected by Swift.
Typical spectral attenuation expected in these sources was summarized by
Mannheim, Hartmann \& Funk (1996), and impacts the sub-100 GeV band when
\teq{z\gtrsim 1}.  Stecker \& de Jager (1996) concluded that the 18 GeV
photon seen by EGRET from GRB 940217 imposed a constraint of
\teq{z\lesssim 2} for this source in order to evade interaction with
intervening background photons.  In the burst context, the higher
redshift pushes the target photons into the near IR and optical
starlight bands, which are better measured at low to moderate redshift
than the classical IR field.   Additional uncertainty is entailed in the
evolution of such backgrounds with redshift during the epoch of rapid
star formation, yet galaxy count data can yield fairly well constrained
models for the evolving background (Kneiske, Mannheim \& Hartmann 2002).
Since the focus here is on internal pair creation in bursts that is
germane to the EGRET/GLAST band, calculations of attenuation by
cosmological backgrounds are beyond the scope of this presentation.

\section{ATTENUATION OF GRB SPECTRA: EVOLUTIONARY EFFECTS}
 \label{sec:evolution_effect}

The GRB spectral forms adopted here have explicitly assumed, for
simplicity, that there is only one radiation component contributing to
the gamma-ray band.  This is supported by the majority of EGRET burst
data (see Table~1 for spectral indices), with the delayed 18 GeV photon
in GRB 940217 providing only a suggestion of a second component.
Evidence for additional components is contingent upon broad spectral
coverage.  Accordingly an interesting possibility has been offered by
the so-called MILAGRITO burst (Atkins et al. 2000; Atkins et al. 2003),
GRB 970417, a BATSE burst that was seen with \teq{\sim 3\sigma}
significance in the TeV band by the water tank \v{C}erenkov detector
MILAGRITO that served as the prototype for MILAGRO. The extrapolation of
the BATSE spectral data up to the TeV band falls over four decades below
the claimed MILAGRITO flux (see Fig.~9 of Atkins et al. 2003), thereby
arguing strongly for the existence of a second component. No other
bursts have been seen by the MILAGRITO/MILAGRO experiment, nor other
TeV-band telescopes, with MILAGRO (Atkins et al. 2005) and MAGIC (Albert
et al. 2006a) providing interesting upper bounds to two bursts.

Multi-component theoretical models exist in the literature (e.g. 
Meszaros \& Rees 1994; Meszaros, Rees \& Papathanassiou 1994; Katz 1994;
Dermer, Chiang \& Mitman 2000), so the relationship between the current
analysis and models with several radiation components cannot be
summarily dismissed.  Inspection of the developments of Section~2
quickly reveals that the pair production optical depth formalism
presented here is entirely valid for multiple component spectra as long
as they use the BATSE band component as the population of target
photons.  The super-GeV spectra may differ tremendously due to the
presence of additional components, however the attenuation factors will
depend only on the known BATSE/EGRET spectral forms.  Accordingly, the
signatures presented here will be widely applicable to burst spectra in
the EGRET/GLAST band, unless additional components emerge below 100 MeV.

In this paper, the analysis is restricted to cases of a single break in
the single component power-law photon spectrum.  This modest limitation
is motivated by expediency, and can be relinquished when more accurate
representations of a GRB source spectrum are desired. Such improvements
can be effected by approximating the spectrum by a piecewise continuous
broken power-law with a sufficient number of components to achieve the
desired accuracy.  Yet such refinements rarely prove necessary if the
specialized form adopted here in Eq.~(\ref{eq:taupp_final}) is tailored
to agree with published Band model fits (Band et al. 1993; Preece et al.
2000).  The reason for this centers on where the image of the target MeV
photons is realized in the absorbed portion of the spectrum.   The
difference between the sharply broken power-law in
Eq.~(\ref{eq:powerlaw}) and the Band spectrum is manifested only near
the break in the MeV band, leading to generally modest underpredictions
of the attenuation factors, and then only near the depths of the troughs
(that appear in
Figures~\ref{fig:GRB930131_spec}--\ref{fig:GRB910503_spec} below) in the
band above 10 GeV.  This is the low-flux portion of the burst spectrum
that will generally fall below current instrumental sensitivities, and
in some cases be obscured by attenuation of cosmological background
radiation fields.  Accordingly the sharp breaks assumed here shall
suffice for the present investigation.

\subsection{Typical Absorption Characteristics}
\label{sec:absorption}

To complete the determination of optical depths, one requires knowledge
of the source emission region size \teq{R} and the number density (or
spectral normalization parameter) \teq{n_\gamma} of photons therein. For
nearby bursts that sample Euclidean space, the coefficient
\teq{n_\gamma} can be obtained from the flux \teq{f_{511}} at 511 keV
per 511 keV energy interval via the simple relation \teq{f_{511}=4\pi
n_{\gamma}cR^2/d^2}, where \teq{d} is the distance to the source.  For
such an absence of cosmological corrections, the perceived source size
\teq{R} can be written as \teq{R=R_v \Gamma^{\lambda}}, where
\teq{\Gamma} is the bulk Lorentz factor of the emission region with
respect to the observer at infinity, and \teq{R_v=c\Delta t} is the
effective size of the region as deduced from observed variability
timescales \teq{\Delta t}.  Following Baring \& Harding (1996, 1997b),
the value of \teq{R_v=3\times 10^7}cm is adopted here, commensurate with
BATSE and COMPTEL variability timescales of \teq{\Delta t=}ms, and
anticipated variability timescales from the GLAST Gamma-Ray Burst
Monitor (GBM).  The index \teq{\lambda} is chosen according to the
presumed geometry and structure of the emission region.  If the
constraining structure scale (e.g. of clumpiness, or between colliding
expanding shells) that defines source variability is along the line of
sight (longitudinal) to the observer, then \teq{\lambda =2}.  This is
the preferred scenario in many GRB models that invoke colliding shells
emanating perhaps from a central compact powerhouse.  When the
variability couples to structure transverse to the line of sight, then
\teq{\lambda =1}, a situation that often appears in models of jet
emission in active galactic nuclei.  Such transverse variability was
adopted by Baring \& Harding (1997a,b).  Theoretical arguments can be
made for both \teq{\lambda=1} and \teq{\lambda=2} scenarios (or hybrids
thereof), though observational discrimination between them is not yet
afforded by the data.   Both cases are considered in this paper,
exhibiting only small qualitative differences between the two.

The earlier analyses of Baring (1993, 1994) and Baring \& Harding (1996,
1997a,b) mentioned cosmological corrections only in passing, and
moreover in an incomplete fashion, primarily because these works
predated BeppoSax burst detections that spawned redshift determinations
via their afterglows.  At this juncture, there is an observational
mandate to include correct redshift dependence in the pair production
analysis.  Such correction factors to the Euclidean (low \teq{z}) case
enter in several ways.  The simplest one connects to the source
variability, whose intrinsic timescale is actually \teq{\Delta t/(1+z)}.
 This is the only redshift contribution to the inferred source size if
the Lorentz factor \teq{\Gamma} is assumed independent of redshift. 
This restriction on \teq{\Gamma} may be inaccurate, since there can be a
significant (and as yet unknown) correlation between the nature of the
burst progenitor and circumburst environment and its redshift.

The major redshift dependence in the problem enters through the
evolution of the source spectrum in its transmission to Earth.  This
receives contributions from the expansion of comoving volumes, which
alters the perceived density of photons in the source, the redshifting
of photon energy, and the frequency with which photons arrive at a
detector.  Spectral transmission calculations essentially connect to a
luminosity transmission formalism, which is textbook and can be found,
for example, in Chapter 5 of Longair (1998).  The core property is that
the comoving, isotropic bolometric source luminosity
\teq{L_{\hbox{\sixrm bol}}} and the detected bolometric {\it energy}
flux \teq{S_{\hbox{\sixrm bol}}} are coupled via the {\it luminosity
distance} \teq{\dlum}:
\begin{equation}
   S_{\hbox{\sixrm bol}} \; =\; \dover{L_{\hbox{\sixrm bol}}}{4\pi \dlum^2}
      \; =\; \dover{R^2}{\dlum^2} \; S_{\hbox{\sixrm GRB,z}} \quad .
 \label{eq:Sbol_Lbol_rel}
\end{equation}
Here \teq{S_{\hbox{\sixrm GRB,z}}=L_{\hbox{\sixrm bol}}/(4\pi R^2)} is
the GRB source flux at its outer periphery, i.e. at redshift \teq{z},
and is a measure intrinsic to the source.  While \teq{L_{\hbox{\sixrm
bol}}} is integrated over solid angles, \teq{S_{\hbox{\sixrm bol}}} is
independent of any collimation, provided the observer lies within the
cone of collimation.

The luminosity distance thus encapsulates the evolution of the energy
fluxes in the curved spacetime of the evolving universe, essentially
coupling to the conservation of energy in the comoving frame. Observe
that both \teq{S_{\hbox{\sixrm bol}}} and \teq{S_{\hbox{\sixrm GRB,z}}}
possess units of energy per unit area per unit time (e.g. erg cm$^{-2}$
sec$^{-1}$). For a particular choice of cosmology, the form for
\teq{\dlum} is
\begin{equation}
   \dlum\; =\;  \dover{c\, (1+z) }{H_0\,\sqrt{\vert 1-\Omega\vert}} \; {\cal S}_k(\Theta )\quad ,
 \label{eq:dlum_def}
\end{equation}
where \teq{H_0} is Hubble's constant, and for 
\teq{k=\hbox{sgn}(\Omega_m+\Omega_{\Lambda}-1)},
\begin{equation}
   {\cal S}_k(\Theta )\; =\; \cases{ \sin\Theta , & \quad $k=1$,\cr
                                                \Theta , & \quad $k=0$, \cr
                                                \sinh\Theta , & \quad $k=-1$,}
 \label{eq:S_k_def}
\end{equation}
is the factor in the Robertson-Walker metric that describes the solid
angle modification from flat space.  Its argument \teq{\Theta} is the 
{\it development angle}, and is given by
\begin{equation}
    \Theta\; =\;  \sqrt{\vert 1-\Omega\vert}  \int_0^z \dover{dz'}{E(z')}\quad ,
    \quad \vert 1-\Omega\vert\; > \; 0\quad ,
 \label{eq:dev_angle_def}
\end{equation}
with \teq{\Omega = \Omega_m+\Omega_{\Lambda}} being the total
density expressed as a fraction of the closure density 
\teq{\rho_c = 3 H_0^2/(8\pi G)}, and 
\begin{equation}
   E(z)\; =\; \sqrt{ \Omega_m (1+z)^3 
    + \Omega_{\Lambda} + (1-\Omega )\, (1+z)^2 }\quad .
 \label{eq:Ez_def}
\end{equation}
Here, the radiation contribution to the universe's dynamics is assumed
negligible, as it is for the epochs since burst genesis.  This form for
the luminosity distance is employed in standard expositions on supernova
probes of cosmology (e.g. Perlmutter, et al. 1997; Riess, et al. 1998).

In keeping with the results of those programs, and following on from the
precise determinations of cosmology by WMAP microwave anisotropy
observations (e.g. Spergel et al. 2003), here it is assumed that
\teq{H_0=0.72}km/sec Mpc$^{-1}$, and a choice of \teq{\Omega_m=0.27} and
\teq{\Omega_{\Lambda}=0.73} is made (i.e. \teq{\Omega\to 1}),
corresponding to a flat space, \teq{k=0} approximation.  This simplifies
the form of \teq{E(z)}, and moreover the expression for \teq{\dlum}. 
Numerical integration of Eq.~(\ref{eq:dev_angle_def}) is routine, and
pertinent asymptotic limits are \teq{\Theta\approx z\, \sqrt{\vert
1-\Omega\vert}} for \teq{z\ll 1}, and \teq{\Theta\to 2 \sqrt{\vert
1-\Omega\vert} /(1-\Omega_{\Lambda})^{1/3}\, F(1/6,\, 2/3;\, 7/6;\,
\Omega_{\Lambda})} for \teq{z\to\infty}. These behaviors guide one to
derive the useful analytic approximation for
\teq{\Omega_{\Lambda}=1-\Omega_m=0.73}:
\begin{equation}
   \dlum\;\approx\; \dover{c}{H_0}\; \dover{z\, (1+z)}{1 + 0.29 z}\;
   \biggl\{ 1 - \dover{z^2}{9(16+z^2)}\,\biggr\}\quad ,
 \label{eq:dlum_approx}
\end{equation}
which is accurate to better than around 2\% for \teq{0 < z < 40},
and generally better than 1\% in the interval \teq{1.7 < z < 40}.

Eq.~(\ref{eq:Sbol_Lbol_rel}) can be easily converted into a spectral
equivalent, and moreover one appropriate for photon number densities, by
expressing the fluxes as integrals over differential photon spectra. 
For general spectral forms \teq{n (\erg)}, the GRB source flux can be
written as
\begin{eqnarray}
   S_{\hbox{\sixrm GRB,z}} &=& 4\pi\, m_ec^3 
                  \int_0^{\infty} \erg_z\, n(\erg_z)\, d\erg_z\nonumber\\[-5.5pt]
 \label{eq:S_GRBz_def}\\[-5.5pt]
         &=& 4\pi\, (1+z)^2\, m_ec^3 
                  \int_0^{\infty} \erg_0\, n\Bigl( [1+z]\, \erg_0\Bigr)\, d\erg_0\quad ,\nonumber
\end{eqnarray}
where \teq{\erg_z = [1+z] \erg_0} is the photon energy at the source
corresponding to an observed dimensionless energy of \teq{\erg_0}.  An
identical form to Eq.~(\ref{eq:S_GRBz_def}) is realized for
\teq{S_{\hbox{\sixrm bol}}}, with the substitution \teq{z\to 0}, so that
the relationship between the integrands of the two flux forms reproduces
Eq.~(7) of Woods \& Loeb (1995).  For the purposes of normalization in
this paper, if an observer detects a differential photon number flux
spectrum \teq{f(\erg_0)=f_{511}\, \erg_0^{-\alpha}} above \teq{\ec}
(with \teq{f_{511}} in units of cm$^{-2}$ sec$^{-1}$, and
\teq{f_{511}/(\alpha -1)} being the total number flux above 511 keV),
assuming \teq{\ec\ll 1} for simplicity, then
\begin{equation}
   S_{\hbox{\sixrm bol}}\; =\; f_{511}m_ec^2 \int_{\ec}^{\infty} \erg_0^{1-\alpha}\, d\erg_0\quad ,
 \label{eq:Sbol_eval}
\end{equation}
a result that is convergent for \teq{\alpha >2}.  Then Eqs.~(\ref{eq:Sbol_Lbol_rel}), 
(\ref{eq:S_GRBz_def}) and~(\ref{eq:Sbol_eval}) can be combined to yield
\begin{equation}
   f_{511}\;\equiv\; (0.511)^{1-\alpha}\, f(1\,\hbox{MeV})
   \; =\; 4\pi\, n_{\gamma}c\, \dover{R^2}{\dlum^2} \; (1+z)^{2-\alpha}\;\; ,
 \label{eq:f511_norm}
\end{equation}
independent of the choice of \teq{\ec}. Inversion of this result then
leads to the specification of \teq{n_{\gamma}} in terms of observables. 
Accordingly, the redshift dependence of the optical depth \teq{\taupp}
is encapsulated entirely in the factor \teq{\dlum^2 (1+z)^{\alpha -2}},
a result that can be inferred from Eq.~(8) of Woods \& Loeb (1995), and
also Eq.~(5.68) of Longair (1998). Note that this form contrasts the
erroneous claim made by Lithwick \& Sari (2001)  that \teq{\taupp\propto
(1+z)^{2\alpha -2}}. In fact, just a single power \teq{\alpha} of
\teq{1+z} should appear since, due to pair creation threshold
kinematics, the optical depth must trace one power of the inverse of the
target (i.e., observed) photon distribution, together with additional
\teq{1+z} factors that are not directly connected to the spectrum.

Note that since \teq{\dlum \approx cz/H_0\equiv d} for \teq{z\ll 1},
Eq.~(\ref{eq:f511_norm}) reduces to the nearby source form
\teq{f_{511}=4\pi n_{\gamma}cR^2/d^2} in the \teq{z\ll 1} limit. In the
implementations here, \teq{\alpha\to\alpha_h} is set. The relationship
between the theoretical flux parameter \teq{f_{511}} and the the
observed flux \teq{f(1\,\hbox{MeV})} at 1 MeV per MeV interval
(tabulated for selected EGRET bursts in Table~1) is included in
Eq.~(\ref{eq:f511_norm}) for future reference. This concludes the
formalism on cosmological input for the pair production attenuation
problem.

The attenuation of a theoretical gamma-ray burst continuum can now be
computed. Neglecting the influence of pair cascading, if the emission
region is comparatively confined, the attenuation of a spectrum like
that in Eq.~(\ref{eq:powerlaw}) is exponential in the optical depth
\teq{\taupp} given in Eq.~(\ref{eq:taupp_final}). However, generally it
can be distributed spatially with skin-depth effects in radiative
transfer operating, so that the attenuation is by a factor of
approximately \teq{1/(1+\taupp )}.  Both cases are exhibited in the
numerical results that follow.  The principal spectral structure
associated with pair production attenuation is illustrated in
Fig.~\ref{fig:GRB930131_spec} for the BATSE/EGRET burst GRB930131, the
so-called Superbowl burst.  This bright source exhibited the flattest
spectrum of all EGRET bursts, and so is the prototype candidate for a
burst easily detectable by GLAST at 100 MeV and above (see Band, et al.
2004, Cohen-Tanugi, et al. 2004, or Omodei et al. 2006, for brief
discussions of expected GLAST performance for burst detections).   Since
GRB 930131 pre-dated the determination of redshifts in optical
afterglows, a redshift of \teq{z=1} was assumed in the modeling (i.e.,
luminosity distance \teq{\dlum =6.54}Gpc), typical of BeppoSax and Swift
long-duration bursts. Relevant source data are listed in Table~1, which
is a composite and extension of Table~1 of Baring \& Harding (1997a) and
Table~2 of Baring \& Harding (1997b). While such flux-selected bursts
might well be at redshifts lower than unity, GRB 990123 at redshift
\teq{z\geq 1.61} (Hjorth et al. 1999) provides a counterexample.

\tableoutpdf{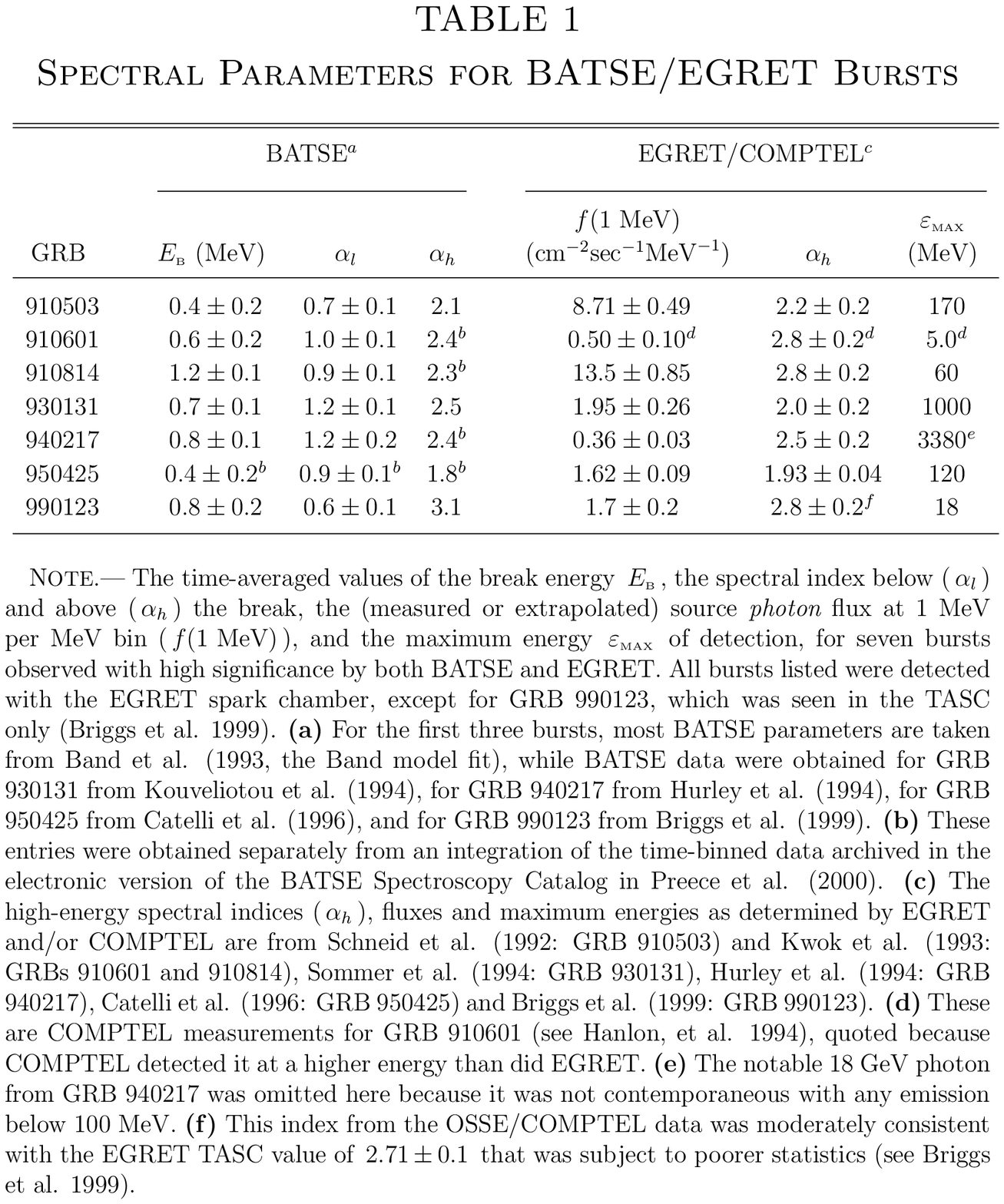}{3.9}{0.0}{0.0}{0.0}

Spectra are presented for different Lorentz factors \teq{\Gamma =
1/\theta_m} (this coupling was adopted by Baring \& Harding 1997a,
though in general other \teq{\Gamma\sim 1/\theta_m} choices are equally
acceptable), incorporating only internal absorption and not treating
propagational attenuation through background radiation fields.  The
spectra indicate marked absorption whose onset energy couples mostly to
the photon flux, source distance and \teq{\Gamma}, as detailed, for
example, in the extensive investigations of Baring (1993) and Baring \&
Harding (1997b), and also depends weakly on the EGRET band spectral
index \teq{\alpha_h}.  For \teq{1/(1+\taupp )} attenuation, above this
turnover, the immediate spectral index is \teq{2\alpha_h-1}, which
flattens to \teq{\alpha_h+\alpha_l-1} at energies higher than
\teq{\Gamma^2/\eb}, where the continuum below \teq{\eb} is sampled as
the pool of target photons.  Such inferences can be made by inspection
of the two terms contributing to Eq.~(\ref{eq:taupp_final}). These
results resemble the character of the spectra illustrated in Baring \&
Harding (1997a) and Baring (2001), though note that higher Lorentz
factors are appropriate here because of the greater luminosity distances
involved (e.g. \teq{\dlum =6.54}Gpc here as opposed to \teq{1}Gpc in
Baring 2001).  Spectra above around \teq{30}GeV must, of course, be
convolved with external absorption due to the cosmic infra-red and
optical backgrounds for typical burst redshifts.

\figureoutpdf{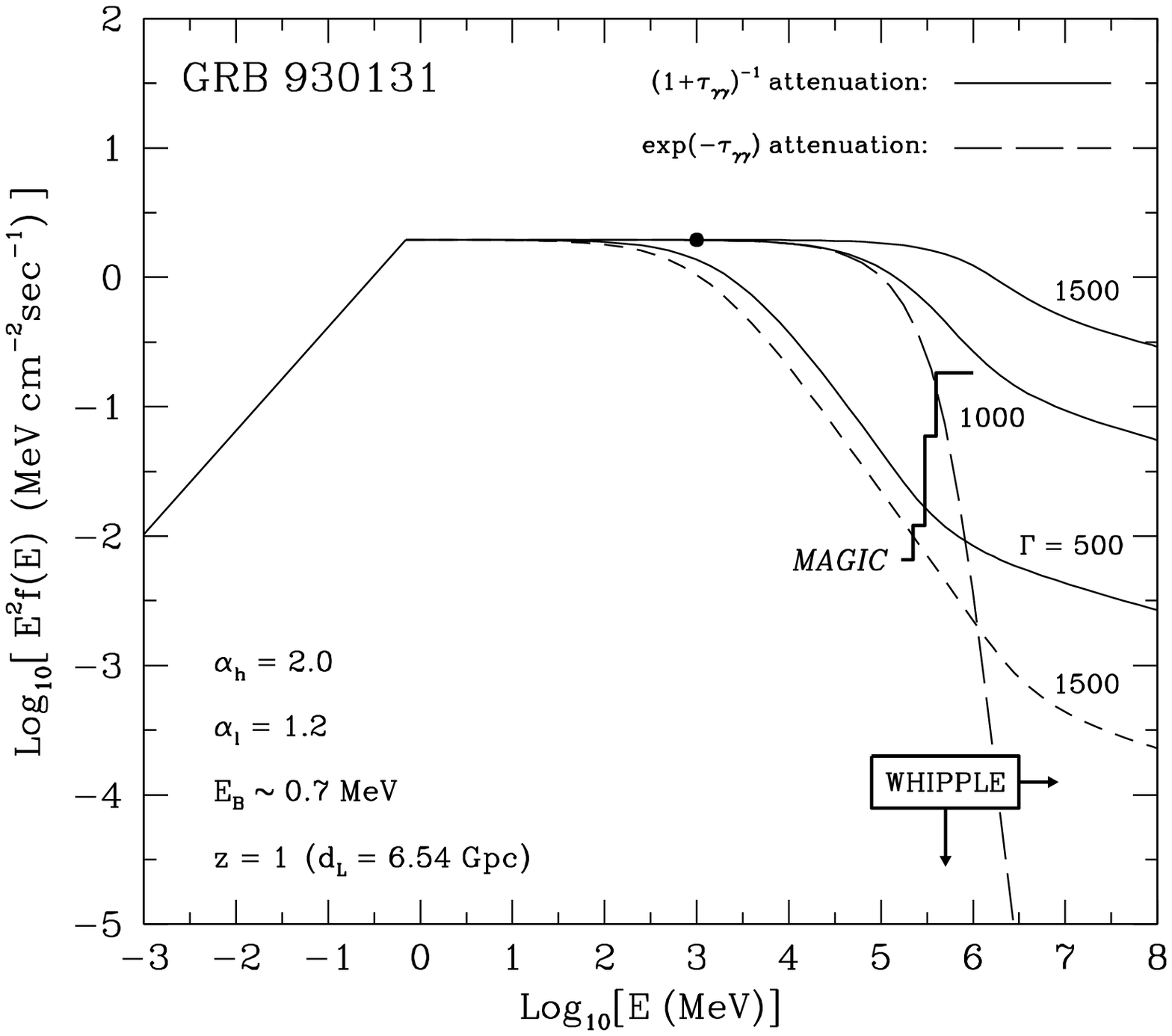}{3.8}{0.0}{-0.2}{
The $\gamma$-$\gamma$ attenuation, internal to the source, for GRB
930131 for bulk Lorentz factors \teq{\Gamma =500,\, 1000,\, 1500} of the
emitting region, assuming a redshift \teq{z=1} typical of long duration
bursts.  The source spectrum (in \teq{\erg^2 f(\erg )} format) was a
power-law broken at \teq{E_{\rm B} =0.7}MeV, with spectral indices
\teq{\alpha_l=1.2} and \teq{\alpha_h=2.0} (see Table~1).  Cases of
\teq{1/(1+\taupp )} attenuation (solid curves) and exponential
attenuation (long dashed curve; \teq{\Gamma =1000} only) are exhibited,
for which a variability size of \teq{R=3\times 10^7\Gamma^2\,\hbox{cm}}
was assumed.  Also, the short dashed line depicts a \teq{\Gamma = 1500},
\teq{1/(1+\taupp )} attenuation example for \teq{R=3\times
10^7\Gamma\,\hbox{cm}}. The filled circle denotes the highest energy
EGRET photon at \teq{\sim 1000} MeV (Sommer et al. 1994).  The threshold
and sensitivity for the post-\teq{T_{90}} Whipple observations of later
bursts (Connaughton et al. 1997) is indicated by the ``WHIPPLE'' box,
while the MAGIC upper bounds for GRB 050713a (Albert et al. 2006a) that
coincided in part with the prompt emission are denoted by the
heavyweight histogram.
 \label{fig:GRB930131_spec}
}

The potential for observational diagnostics is immediately apparent.
First, the extant EGRET data already provides a lower bound to
\teq{\Gamma}: the dot on Fig.~\ref{fig:GRB930131_spec} represents the
highest energy photon from GRB930131, and clearly suggests that
\teq{\Gamma\gtrsim 700} if the burst was at a redshift of \teq{z\gtrsim
1}.  It is also clear from the figure that the exponential and
\teq{1/(1+\taupp )} attenuation signatures are strikingly different, so
that a sensitive hard gamma-ray detector will easily be able to
discriminate between these two possibilities. While realization of the
\teq{1/(1+\taupp )} case is anticipated, should exponential attenuation
be observed, it would be a profound diagnostic indicating that the
emission region is relatively small, confined to a scale length shorter
than the mean free path for pair attenuation.  Comparison of the
\teq{R=3\times 10^7\Gamma^2\,\hbox{cm}} and \teq{R=3\times
10^7\Gamma\,\hbox{cm}} examples for \teq{\Gamma =1500} in
Fig.~\ref{fig:GRB930131_spec} indicates that, in the absence of
time-varying spectra, it will be difficult to discriminate between these
longitudinal and transverse variability cases, since they cannot be
clearly resolved from Lorentz factor inferences.  Specifically,
\teq{R\propto \Gamma} scenarios require slightly higher \teq{\Gamma} for
transparency at a given photon energy than do longitudinal variability
cases.

To aid in the interpretation of pair attenuation it is useful to
analytically determine the mean interacting (target) photon energy
\teq{\erg_{\hbox{\sixrm int}}(\erg)} for a given observed photon energy
\teq{\erg}.  This is computed as the ratio of an interaction probability
like that in Eq.~(\ref{eq:taupp}), but with the extra weighting factor
\teq{\omega} in the integrand, to the interaction probability with unit
weighting factor.  Retracing the analytic development in
Sec.~\ref{sec:gen_form}, it is quickly deduced, using the spectral form
in Eq.~(\ref{eq:ndef}) for the target photon population, that
\begin{equation}
   \erg_{\hbox{\sixrm int}}(\erg)\;\sim\; \dover{4}{\erg\theta_m^2}\,
      \dover{{\cal K}(\alpha -1,\, \etac\theta_m)}{{\cal K}(\alpha ,\, \etac\theta_m)}\quad .
 \label{eq:erg_int_calc}
\end{equation}
For this spectral specialization, \teq{\ec} plays the approximate role
of the break energy \teq{\eb}, though an alternative and slightly more
involved form can easily be obtained to pertain to the spectrum in
Eq.~(\ref{eq:powerlaw}). Ratios of \teq{{\cal K}} functions can be
inferred from Fig.~\ref{fig:calK_char}, and are typically of the order
of a few for \teq{\eta\sim 1}.  Accordingly, for the specific example in
Fig.~\ref{fig:GRB930131_spec}, Eq.~(\ref{eq:erg_int_calc}) clearly
establishes that the attenuation image of the BATSE-band spectral break
at \teq{\eb} appears at the energy \teq{\erg \sim 4\Gamma^2/\eb}, which
corresponds to the pair threshold condition in the bulk fluid rest
frame, i.e., \teq{ (\erg/\Gamma )\, (\eb/\Gamma )\gtrsim 4}.  This is
generally at or above the threshold energy of atmospheric \v{C}erenkov
telescopes.

The first intensive TeV-band exploration of bursts was the Whipple
Telescope campaign (Connaughton et al. 1997), whose results are typified
by the ``WHIPPLE'' box in Fig.~\ref{fig:GRB930131_spec}. This campaign
in 1994-1995 consisted of targeted observations of nine BATSE bursts
within, at best, around two minutes of trigger, coordinated via the
BACODINE network.  For each burst, the Whipple observations comprised 6
scans of around 30 minutes each with slightly offset sky positions (by
\teq{\approx 3^{\circ}}) to cover the BATSE error boxes.  No TeV-band
detections were made, and so the WHIPPLE box in
Fig.~\ref{fig:GRB930131_spec} represents upper limits to the potential
flux from the bursts studied in Connaughton et al. (1997).  Due to the
slewing times of a few minutes, these bounds are more germane to the
burst afterglow phase, for which the MeV-band flux level is unkown. In
contrast, the recent rapid-slewing observation (Albert et al. 2006a) of
GRB 050713a by the MAGIC Telescope, triggered by a Swift alert, enabled
a viewing of the burst at about 40 seconds after its onset, for a period
of 37 minutes (Swift determined \teq{T_{90}=70\pm 10}sec). No gamma-ray
signal was detected by MAGIC, and the resulting upper bounds are
indicated in Fig.~\ref{fig:GRB930131_spec} as the heavyweight histogram.
Since no redshift was determined for GRB 050713a, this sensitivity
histogram corresponds to the \teq{z=1} flux upper limit listed in
Table~1 of Albert et al. (2006a).  This flux limit is approximately
commensurate with the fluence limits obtained by the MILAGRO
\v{C}erenkov water tank experiment for GRB 010921 (Atkins et al. 2005),
the only burst surveyed in the MILAGRO campaign of 2000-2001 that
possessed a moderately low, measured redshift (\teq{z=0.45}).  STACEE
also obtained a flux upper limit to GRB 050607 (Jarvis et al. 2005). 
The MAGIC sensitivity depicted in Fig.~\ref{fig:GRB930131_spec} is
representative of the present capability of state-of-the-art atmospheric
\v{C}erenkov systems in terms of prompt GRB observations.

\subsection{Time Evolution of Spectra}
\label{sec:evolution}

The photon counting statistics associated with EGRET burst detections
generally only permitted using time-integrated spectra for the purposes
of pair attenuation considerations.  The enhanced sensitivity offered by
GLAST by late 2007 may permit some exploration of time-dependent
attenuation characteristics that are previewed in this Section,
particularly if they emerge in bright bursts in the energy band below
around 1 GeV.

There are a plethora of time-dependent gamma-ray burst spectral models
in the literature, often pertaining to afterglow evolution.  For the
prompt phase, there is evidence of prevalent hard-to-soft evolution on
average in the BATSE database, though the spectral hardness development
during a burst can often be more complicated (e.g. Ford et al. 1995,
Preece et al. 1995).  If the Lorentz factor \teq{\Gamma} is effectively
constant, then this observed variation could be due to evolving
dissipational characteristics in the internal shocks, which are subtle
and difficult to encapsulate in a cohesive manner.  As an alternative
scenario, the spectral evolution could primarily reflect the change in
\teq{\Gamma} on a timescale of the burst duration.  For long-duration
bursts with \teq{T_{90}\sim 1}--\teq{100}sec, the physical scale
associated with their duration is \teq{\Gamma^2 c\, T_{90} \sim 3\times
10^{15}}--\teq{3\times 10^{18}}cm for \teq{\Gamma\sim 300}--\teq{1000}.
Hence bulk deceleration during the burst prompt phase is likely, and is
a natural driver for GRB dissipation (e.g., Rees and M\'esz\'aros 1992).
Both evolutionary possibilities are examined here.

The focus is first on a constant \teq{\Gamma}, constant \teq{\alpha_h}
scenario with the flux allowed to vary, for which representative spectra
are displayed in Fig.~\ref{fig:GRB950425_spec} for the case of GRB
950425.   There, the emission spectrum evolves only by an
across-the-board decline of flux with time, displayed in increments of
half a decade per ``snapshot'' (solid curves). Accordingly, this case
does not explore adiabatic deceleration issues, or the role of radiative
cooling, but does provide a definite correlation between evolution of
the BATSE-band flux and the turnover energy \teq{E_t} in the GLAST
window.  To elucidate this interesting diagnostic, observe that since
the Lorentz factor is held constant, and at the turnover energy the last
term in Eq.~(\ref{eq:taupp_final}) provides the dominant contribution to
\teq{\taupp}, one finds that \teq{\taupp\propto f(\hbox{1 MeV})\,
E_t^{\alpha_h-1}}.  From this, it follows that the turnover energy
couples to the MeV-band flux according to
\begin{equation}
   E_t\;\propto\; [\, f(\hbox{1 MeV})\, ]^{-1/(\alpha_h-1)}\quad ,
 \label{eq:eturn_form}
\end{equation}
established using a \teq{\taupp\sim 1} criterion.  For the GRB 950425
example in the Figure, this gives \teq{E_t\;\sim\; [\, f(\hbox{1 MeV})\, ]^{-1/2}}.
Combining this result with the \teq{f(\hbox{1 MeV})\, E_t^{2-\alpha_h}}
spectral form far above the break energy \teq{E_{\rm B}} then quickly yields
\begin{equation}
   E_t^2\, f(E_t)\;\propto\; E_t^{3-2\alpha_h}
       \;\propto\; [\, f(\hbox{1 MeV})\, ]^{(2\alpha_h-3)/(\alpha_h-1)}
 \label{eq:eturn_locus}
\end{equation}
as the mathematical form for the locus of turnover points in \teq{\nu
F_{\nu}} space.  For GRB 950425 this generates
\teq{E_t^{3-2\alpha_h}=E_t^{-0.92}}, since  \teq{\alpha_h=1.93}, which
is represented by the dashed curve in Fig.~\ref{fig:GRB950425_spec}. 
Clearly, this correlation is extremely pronounced, and is a very useful
probe of the constancy of \teq{\Gamma} if the high energy spectral index
does not evolve. In bright, flat-spectrum sources such as GRB 950425,
GLAST may well be suited to investigating whether such character is
exhibited, provided that the turnovers appear at below 1 GeV so that the
LAT can accumulate enough photons in a typical burst duration. In
particular, for the \teq{\Gamma} adopted in this example, clearly the
relative probability of seeing a super-100 MeV photon to observing a 1
MeV photon is enhanced at later times; whether this is connected to the
detection (see Hurley et al. 1994) of the 18 GeV photon in GRB 940217
over an hour after burst trigger is undetermined. Note also that the
attenuation results apply to any time-ordering of the curves, not just
monotonically declining fluxes. Accordingly, one would infer a decline
of the turnover energy in time in a rising flux precursor to the prompt
GRB emission.

Another property of the attenuation is that well above the turnover
energy, the results are approximately independent of \teq{f(\hbox{1
MeV})}, when \teq{\eb} is held fixed in time.  This arises in the
``recovery domain,'' where the target photons are below \teq{\eb} and
the optical depth \teq{\taupp} is declining with energy till eventually
the internal attenuation becomes insignificant (illustrated in Baring \&
Harding 1997b).  At such energies higher than \teq{\Gamma^2/\eb}, as
\teq{\alpha_l=0.9} for GRB 950425, the optical depth is almost
independent of the photon energy so that it mostly scales as
\teq{f(\hbox{1 MeV})}.  This behavior almost exactly cancels the
normalization of the emission continuum so that the emergent spectrum is
virtually independent of the MeV-band flux.  For this source, given the
location of the ACT sensitivity indicator in Fig.~\ref{fig:GRB950425_spec} 
(i.e. MAGIC flux upper limits for GRB 050713a), and the probable role of 
additional line-of-sight attenuation due to cosmic background radiation 
fields, the recovery domain is unlikely to be accessible to current 
generation \v{C}erenkov telescopes.

\figureoutpdf{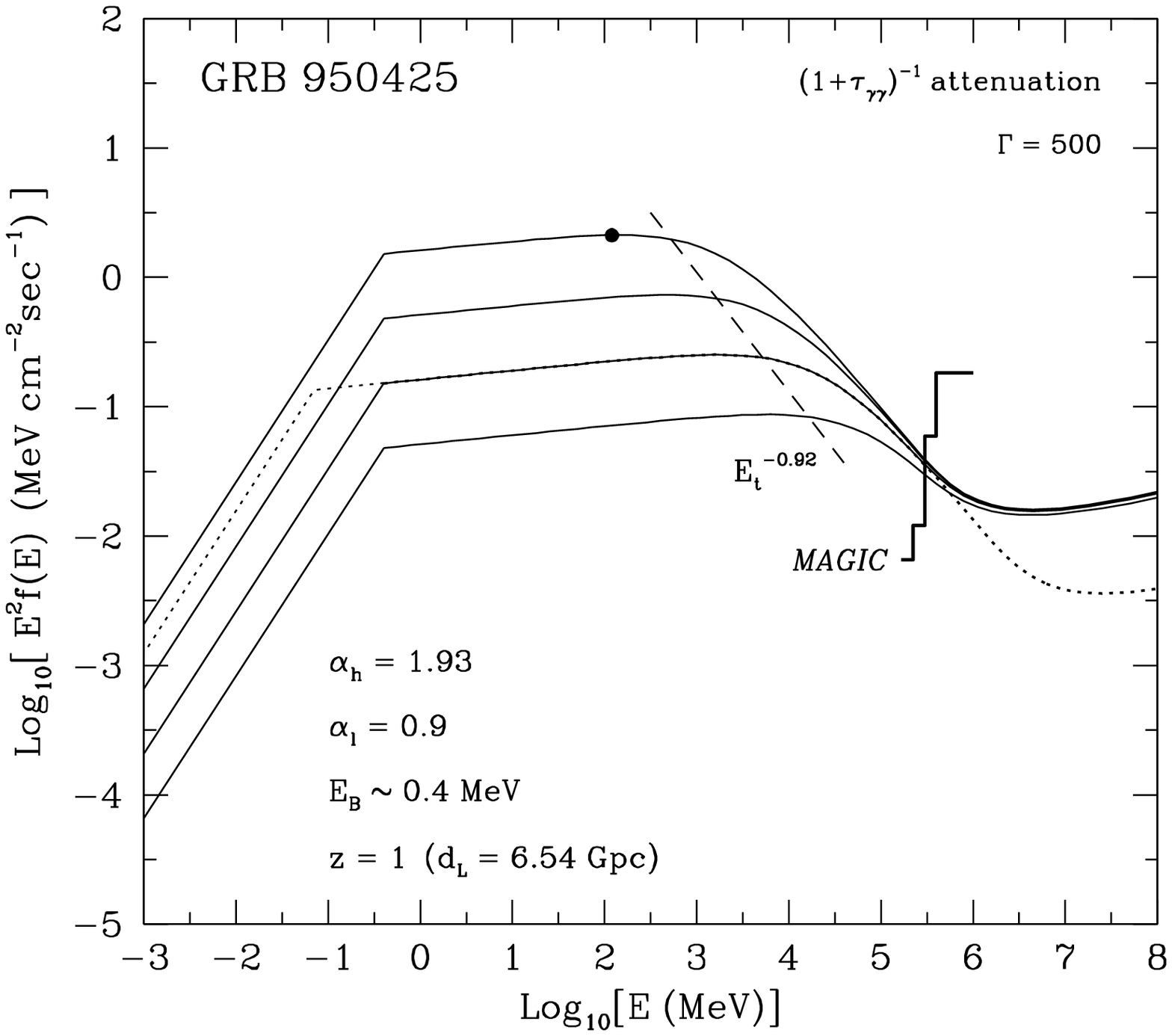}{3.8}{0.0}{-0.2}{
The $\gamma$-$\gamma$ attenuation, internal to the source, for GRB
950425 for bulk Lorentz factor \teq{\Gamma =500}, assuming a redshift
\teq{z=1} typical of long duration bursts.  The source spectrum (in
\teq{\erg^2 f(\erg )} format) was a power-law broken at \teq{E_{\rm B}
=0.4}MeV (solid curves), with spectral indices \teq{\alpha_l=0.9} and
\teq{\alpha_h=1.93} (see Table~1).  Only \teq{1/(1+\taupp )} attenuation
is depicted, and a variability size of \teq{R=3\times
10^7\Gamma^2\,\hbox{cm}} was assumed.   The peak flux evolution between
the curves is by successive factors of \teq{\sqrt{10}} from top to
bottom.  The dotted curve illustrates a hard-to-soft evolution situation
and is virtually coincident with the corresponding fixed \teq{E_{\rm B}
=0.4}MeV case above the break, and below 500 GeV. The filled circle
denotes the highest energy EGRET photon at \teq{\sim 120} MeV (see
Table~1). The attenuation turnover energy \teq{E_t} evolves, tracing out
an \teq{E_t^{-0.92}} locus, as discussed in the text. As in
Fig.~\ref{fig:GRB930131_spec}, the MAGIC flux limit upper bounds for GRB
050713a (Albert et al. 2006a) are denoted by the heavyweight histogram.
 \label{fig:GRB950425_spec}
}

Finally, for this example, observe that since \teq{\eb} is not
influential in determining the turnover energy, it was held constant for
illustrative purposes.  In Fig.~\ref{fig:GRB950425_spec}, the dotted
curve depicts a case where strong hard-to-soft evolution would arise
subsequent to the peak flux snapshot. Well above \teq{\eb}, and below
the recovery domain at \teq{E\gtrsim 1}TeV, this curve is not
distinguishable from the corresponding solid curve for the fixed
\teq{\eb} case, since the target energies for photons near the turnover
are above the break energy.  In the \teq{E\gtrsim 1}TeV portion of the
spectrum, the extra supply of target photons below \teq{0.4} MeV
relative to the \teq{\eb =0.4}MeV solid line case with the same
\teq{f(\hbox{1 MeV})} clearly enhances the pair opacity and lowers the
spectrum accordingly. Fig.~\ref{fig:GRB950425_spec} encapsulates high
energy spectral attenuation characteristics appropriate for constant
\teq{\Gamma} scenarios that admit evolution of the break energy
\teq{\eb} with time, and therefore possesses broader applicability.  The
import of this broad scope will soon become apparent, when scenarios
incorporating \teq{\Gamma} evolution with time are addressed just below.
Observe also that while not illustrated in this example, evolution of
\teq{\alpha_h} with time can be easily accommodated to provide useful
diagnostics on \teq{\Gamma}.

The simplest (but by no means uniquely preferred) case corresponding to
appreciable reductions of \teq{\Gamma} on a timescale of the burst
duration is defined by the hydrodynamic sweep-up by a blast wave of
exterior material from the interstellar medium or a progenitor wind.
This is essentially the scenario explored by Rees and M\'esz\'aros
(1992), and encapsulated nicely in the paper by Dermer, Chiang \&
B\"ottcher (1999). It is this latter analysis that is employed here,
specifically the self-similar solution for blast wave deceleration in a
uniform or spatially-structured circumburst environment.  Assuming, for
simplicity, that the density of material the blast wave impacts is
uniform, then the sweep-up is precisely a relativistic extension of the
classic Sedov-Taylor phase solution for supernova remnant expansion.
This hydrodynamic problem for an expansion with an ultra-relativistic
equation of state was treated by Blandford \& McKee (1976). In the limit
that the magnetic field plays a passive role, the conservation of energy
and momentum in an adiabatically-expanding, non-radiative scenario
yields a radial form for the deceleration of
\begin{equation}
   \Gamma (r)\;\propto\; r^{-3/2}\;\propto\; t^{-3/8}
     \quad , \quad \tdyn\;\ll\; t \;\ll\; \trad
 \label{eq:Gamma_evolve_dyn}
\end{equation}
after the coasting phase has passed on a timescale of \teq{\tdyn}.  
If the expansion is radiative, i.e. loses energy via radiation on times 
\teq{\trad} shorter than dynamical timescales \teq{\tdyn}, then the 
evolution is described by a momentum-conserving ``snowplow'' form 
(e.g. Blandford \& McKee 1976; Dermer, Chiang \& B\"ottcher 1999) 
\begin{equation}
   \Gamma (r)\;\propto\; r^{-3}\;\propto\; t^{-3/7}
     \quad , \quad \trad\;\ll\; \tdyn \;\ll\; t\quad .
 \label{eq:Gamma_evolve_rad}
\end{equation}
The dynamical sweep-up scale \teq{c\tdyn} is directly expressible 
in terms of the ``explosion'' energy \teq{{\cal E}_0}, the solid angle 
\teq{\Omega_0} of collimated expansion, coasting phase (initial) 
Lorentz factor \teq{\Gamma_0}, and mass density \teq{\rho} of the 
circumburst medium as
\begin{equation}
   c\tdyn\; =\; \dover{1}{\Gamma_0^2}\;
         \biggl\{ \dover{3 {\cal E}_0}{\Omega_0\Gamma_0^2\rho c^2} \biggl\}^{1/3}\quad .
 \label{eq:tdyn}
\end{equation}
Here the presence of the \teq{\Gamma_0^2} factor out the front accounts
for relativistic time dilation between the blast and observer's frames
(similar factors mediate the \teq{r}-\teq{t} coupling in
Eqs.~[\ref{eq:Gamma_evolve_dyn}] and~[\ref{eq:Gamma_evolve_rad}]), and
the other \teq{\Gamma_0^2} factor couples to the energy conservation
during sweep-up; these factors provide the only relativistic
modification from the classic Sedov-Taylor phase timescale.  Other
radial/temporal dependences for \teq{\Gamma} can be envisaged (e.g.
M\'esz\'aros, Rees \& Wijers 1998), for example that incurred by a
power-law radial dependence for the circumburst density profile.

The forms in Eqs.~(\ref{eq:Gamma_evolve_dyn})
and~(\ref{eq:Gamma_evolve_rad}) suffice for the purposes of this
exposition, yielding time-dependence templates for scoping out pair
creation attenuation signatures.  Collimation of the outflow causally
impacts the kinematic phase space only very late in the GRB afterglow. 
The other ingredients needed are the Lorentz factor dependences for the
spectral break energy \teq{\eb} and the flux at this or some fixed
higher energy.   These quantities are model-dependent.  Adhering to the
prevailing paradigm that the GRB emission mechanism is synchrotron
radiation, Dermer, Chiang \& B\"ottcher (1999) argued in their external
shock model that the break energy, if it corresponds to synchrotron
photons from electrons of the minimum Lorentz factor \teq{\gammin} in a
non-thermal distribution, would behave according to
\begin{equation}
   \eb\;\propto\; [\Gamma(r)]^4\quad .
 \label{eq:eb_evolve}
\end{equation}
The origin of this form assumes that \teq{\gammin\propto\Gamma} arises
during energy transfer in the sweep-up phase, that the ambient magnetic
field is boosted by \teq{\Gamma} in the comoving frame, and also
incorporates one power of \teq{\Gamma} for the energy blueshift to the
observer's frame.  All these are reasonable contentions.  Yet, it is
clear that a sizeable fraction of bursts cannot generate low energy
spectra consistent with the synchrotron mechanism using a cutoff
distribution; this is the so-called ``line of death'' issue raised by
Preece et al. (1998, 2000), discussed also in Baring \& Braby (2004). 
Hence, the possibility of  \teq{\eb (\Gamma)} dependences other than in
Eq.~(\ref{eq:eb_evolve}) can also be entertained.

The flux evolution with time or Lorentz factor is encapsulated in the
third part of Eq.~(20) of Dermer, Chiang \& B\"ottcher (1999).
Physically, this form assumes that there is prompt dissipation of a
sizeable fraction the available energy, namely the energy of matter
swept up in the deceleration phase.  For environments where the the
circumburst medium is uniform, this form yields a temporal dependence
for \teq{\erg\gg\eb} of
\begin{equation}
   n(\erg )\; \propto\; \erg^{-\alpha_h}\; t^{-3\varphi /(8-\eta_r)}
   \;\propto\; \erg^{-\alpha_h}\;\Gamma^{\varphi}
 \label{eq:flux_evolve}
\end{equation}
where
\begin{equation}
   \varphi\; =\; 4\alpha_h - \dover{16}{3} + \dover{2}{3}\,\eta_r
 \label{eq:varphi_def}
\end{equation}
is the index defining the bulk Lorentz factor dependence for the source 
flux, and
\begin{equation}
   \eta_r\; =\; \cases{ 0 \vphantom{\bigl(} \;\; , & adiabatic,\cr
                           1 \vphantom{\bigl(}\;\; , & radiative\cr}
 \label{eq:etar_def}
\end{equation}
is the parameter that demarcates radiative and adiabatic expansion
cases. Observe that with this definition, \teq{\Gamma\propto
t^{-3/(8-\eta_r)}} describes the cases in
Eqs.~(\ref{eq:Gamma_evolve_dyn}) and~(\ref{eq:Gamma_evolve_rad}), so
that the identity \teq{g= 3/(2-\eta_r)} provides a connection to the
parameter \teq{g} that Dermer, Chiang \& B\"ottcher (1999) use
throughout their investigation, with \teq{\Gamma\propto r^{-g}}. 
Observe also that inserting Eq.~(\ref{eq:eb_evolve}) into
Eq.~(\ref{eq:flux_evolve}) yields
\begin{equation}
   \eb^2\, n(\eb )\;\propto\; t^{-1-3\eta_r/(8-\eta_r)}
      \;\propto\; \Gamma^{2(4+\eta_r)/3}\quad ,
 \label{eq:peakflux_evolve}
\end{equation}
commensurate with the evolution of the {\it peak flux} that is given in
Eq.~(16) of Dermer, Chiang \& B\"ottcher (1999).  Adiabatic expansions
generate \teq{\eb^2\, n(\eb )\propto t^{-1}\propto \Gamma^{8/3}}, a form
dictated by energy/momentum conservation in sweep-up, which because of
scale-invariance is deducible directly from Eq.~(\ref{eq:tdyn});
radiative expansions produce \teq{\eb^2\, n(\eb )\propto
t^{-10/7}\propto \Gamma^{10/3}}. The flux evolution prescribed by
Eq.~(\ref{eq:flux_evolve}) can be easily modified to accommodate
expansions into non-uniform circumburst media, yet suffices for the
illustrative purposes of this paper.

It should be noted that while the temporal evolution employed in Dermer,
Chiang \& B\"ottcher (1999) envisages an external shock model for the
prompt emission, and can directly be applied to afterglow
considerations, it is also germane for internal shock models if the
gamma-ray activity arises during a deceleration epoch.   Then, the
colliding shocks would possess different Lorentz factors, which would on
average slow down with the expansion.  The relevant \teq{\Gamma} for the
purposes of this analysis would then be some average of those of the
colliding shells, and would be controlled by the hydrodynamics discussed
just above.

\figureoutpdf{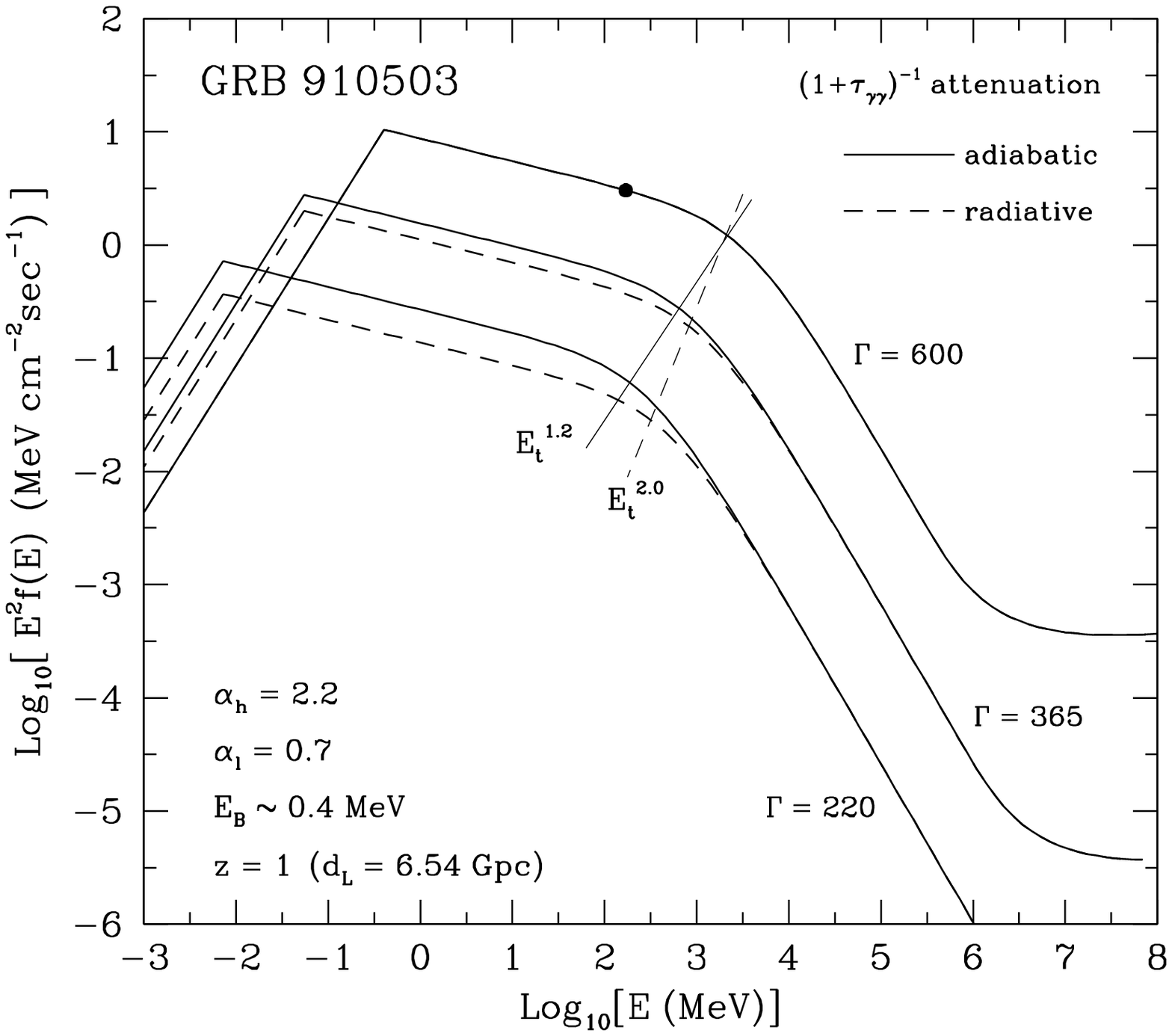}{3.8}{0.0}{-0.2}{
An evolving sequence depicting $\gamma$-$\gamma$ attenuation, internal
to the source, for GRB 910503, again assuming a redshift \teq{z=1}
typical of long duration bursts.  The source spectrum (in \teq{\erg^2
f(\erg )} format) was a power-law broken initially at \teq{E_{\rm B}
=0.4}MeV (uppermost solid and dashed curves that are coincident), with
spectral indices \teq{\alpha_l=0.7} and \teq{\alpha_h=2.2} (see Table~1)
that are maintained throughout. For this initial case, the bulk Lorentz
factor was \teq{\Gamma=600}. The highest energy EGRET photon at
\teq{\sim 170} MeV (Schneid et al. 1992) is denoted by the filled
circle.  Again only \teq{1/(1+\taupp )} attenuation is depicted, and a
variability size of \teq{R=3\times 10^7\Gamma^2\,\hbox{cm}} was assumed.
The peak flux and hard-to-soft evolution between the curves is
indicated via declining Lorentz factors, i.e., cases \teq{\Gamma =365}
and \teq{\Gamma=220} as labelled.  The hard-to-soft evolution of the
break energy \teq{\eb} is prescribed via Eq.~(\ref{eq:eb_evolve}), while
the flux evolution is given in Eq.~(\ref{eq:flux_evolve}), and is
discussed in the text. Solid curves correspond to adiabatic blast wave
deceleration, for which the attenuation turnover energy approximately
traces out [see Eq.~(\ref{eq:eturn_locus_evolve})] the depicted
\teq{E_t^{1.2}} locus, while dashed curves illustrate a strongly
radiative case, for which the locus of \teq{\taupp =1} turnover points
is the illustrated \teq{E_t^{2.0}} line.
 \label{fig:GRB910503_spec}
}

Representative spectra for evolving \teq{\Gamma} scenarios are exhibited
in Fig.~\ref{fig:GRB910503_spec} for the case of GRB 910503.  This
elaborates upon the evolutionary sequence addressed in Baring (2001) for
GRB 930131. Here \teq{\alpha_l} and \teq{\alpha_h} are held constant, so
as to simply demonstrate the key characteristics, though for some bursts
(e.g. see Gonzalez et al. 2004) there are significant variations of
\teq{\alpha_h} with time. The flux and break energy evolve according to
Eqs.~(\ref{eq:eb_evolve}) and~(\ref{eq:flux_evolve}) respectively, and
in this example \teq{\ec=0}. Notably, the turnover energies are now
declining functions of time, contrasting the constant \teq{\Gamma}
example in Fig.~\ref{fig:GRB950425_spec}, for which the turnover locus
is \teq{E_t^{3-2\alpha_h}}.  This behavior is obviously caused by the
increased opacity due to the decline of \teq{\Gamma}, which more than
offsets the tendency of a reduced flux at typical super-MeV target
photon energies to reduce the pair opacity.  Such a signature renders it
easy to discriminate GRB environs where bulk deceleration is prevalent
during the prompt phase, from the case of  constant \teq{\Gamma}, where
internal shock dissipation is completely decoupled from external shock
evolutionary dynamics.  This is a powerful observational diagnostic that
is only modified in detail, but not in character, by model nuances such
as spatial non-uniformity in the circumburst medium.

A comparison of the solid and dashed curves in
Fig.~\ref{fig:GRB910503_spec} indicates that it will be harder to
discern between adiabatic and radiative evolutionary scenarios, though
it may be marginally possible. The tracks of the spectrum around the
break energy do not distinguish substantially between the \teq{\eta_r=0}
and \teq{\eta_r=1} cases in that the separation of fluxes for given
break energy is only modest, and break smoothing could render such
trends indistinguishable.  The turnover feature adds extra information
that is helpful, provided that the hard gamma-ray continuum is detected
with sufficient significance over around at least 2 decades of flux
evolution at above 1 GeV. The Figure marks two loci that the turnovers
trace in time for the respective adiabatic and radiative cases.  As for
the constant \teq{\Gamma} case, these loci define the correlation
between evolution of the BATSE-band flux and the turnover energy
\teq{E_t}, which can again be found using the last term in
Eq.~(\ref{eq:taupp_final}) as the dominant contribution to \teq{\taupp}.
One finds that \teq{\taupp\propto f(\hbox{1 MeV})\,
E_t^{\alpha_h-1}/\Gamma^{\lambda + 2\alpha_h}}, where \teq{\lambda =1,2}
is the index defining the Lorentz contraction relating the perceived
source size to the variability timescale.  From this, and
Eq.~(\ref{eq:flux_evolve}), it follows that the turnover energy
satisfies
\begin{equation}
   E_t\;\propto\; \Gamma^{(\lambda + 2 \alpha_h-\varphi )/(\alpha_h-1)}\quad ,
 \label{eq:eturn_evolve}
\end{equation}
expressing the \teq{\taupp\sim 1} criterion, with \teq{\varphi} as 
defined in Eq.~(\ref{eq:varphi_def}).  Inverting this result and 
combining it with Eq.~(\ref{eq:flux_evolve}), the time-dependent
spectral form valid far above the break energy \teq{\eb}, results in
\begin{equation}
   E_t^2\, f(E_t)\;\propto\; E_t^{\varrho}\quad ,\quad
   \varrho\; =\; 2-\alpha_h + \dover{\varphi (\alpha_h-1)}{\lambda + 2 \alpha_h-\varphi}
 \label{eq:eturn_locus_evolve}
\end{equation}
as the mathematical form for the locus of turnover points in \teq{\nu
F_{\nu}} space.  For GRB 910503 with \teq{\alpha_h=2.2}, and assuming
\teq{\lambda =2}, this approximately generates \teq{E_t^{1.2}} for the
adiabatic case (\teq{\eta_r=0}) and \teq{E_t^{2.0}} for radiative
regimes (\teq{\eta_r=1}), which are represented by the lightweight solid
and dashed lines in Fig.~\ref{fig:GRB910503_spec}.  These are clearly
distinct, and offer the attractive possibility of discriminating between
adiabatic and radiative evolutionary scenarios for a given burst,
particularly if explored in conjunction with the \teq{\eb} evolution.

Since \teq{\varrho} is a rapidly increasing function of \teq{\alpha_h},
flat-spectrum bursts with high super-100 MeV fluxes offer the most
promising discrimination: \teq{\alpha_h=2} yields \teq{\varrho =4/5} for
adiabatic (\teq{\eta_r=0}) regimes, and \teq{\varrho =5/4} for radiative
ones (\teq{\eta_r=1}).  However, note that when adopting a variability
size of \teq{R=3\times 10^7\Gamma\,\hbox{cm}} (i.e., \teq{\lambda =1}),
the loci in Fig.~\ref{fig:GRB910503_spec} change significantly, to
approximately \teq{E_t^{1.95}} and \teq{E_t^{3.7}} for adiabatic and
radiative cases, respectively.  Hence, in practice it may prove
difficult to cleanly distinguish between the effects of non-adiabaticity
in the expansion and the nature of structure in the emission region
using the turnover energy as a diagnostic.  Using the sub-MeV band
temporal traces of the break energy in conjunction should help separate
these competing influences.

\section{THE POTENTIAL FOR OBSERVATIONAL DIAGNOSTICS}
 \label{sec:diagnostics}

As a benchmark for the GRB phase space that is germane to internal
\teq{\gamma\gamma} opacity diagnostics by GLAST and perhaps atmospheric
\v{C}erenkov telescopes, and future experiments down the line, one can
map the contours the turnover energies trace for burst parameters
typical of the bright sources EGRET that detected. For given
\teq{\Gamma} and \teq{\alpha_h}, the maximum photon energy \teq{\Emax
=E_t} (in MeV) attainable in the EGRET-band power-law, before pair
attenuation turnovers appear, is simply determined by noting that the
spectral structure at and below the BATSE-band break is immaterial to
its evaluation.  Therefore, one can set \teq{\eb\to 0} in the analysis,
and then impose a \teq{\taupp \sim 1} criterion to establish, using
Eqs.~(\ref{eq:tauppform}) and~(\ref{eq:calT_beam}), that
\begin{equation}
   \Emax\; =\; \dover{0.511\; \Gamma^{2\alpha_h/(\alpha_h-1)} }{
             \Bigl\{ n_{\gamma}\sigt R\;  {\cal A}(\alpha_h )\, 
                    {\cal H}(\alpha_h,\, 1)  \Bigr\}^{1/(\alpha_h-1)} }\;\;\hbox{MeV}\quad .
 \label{eq:Emax_GLAST}
\end{equation}
This energy is plotted as a function of \teq{\alpha_h} for different
\teq{\Gamma} and fluxes \teq{f(\hbox{1 MeV})} in
Fig.~\ref{fig:Emax_phasespace}, specifically for source luminosity
distances of 6.54 Gpc (i.e. \teq{z=1}). Such curves were determined
assuming a variability size of \teq{R=3\times 10^7\Gamma^2\,\hbox{cm}},
and are applicable for both \teq{1/(1+\taupp )} attenuation and
exponential attenuation. This depiction represents an update of Figure~6
of Baring and Harding (1997b) that reflects the moderate redshifts now
associated with GRBs in the Swift/HETE era; Jakobsson et al. (2006)
report a mean redshift of \teq{z\sim 2.8} in their compilation of 16
Swift bursts with measured redshifts.

\figureoutpdf{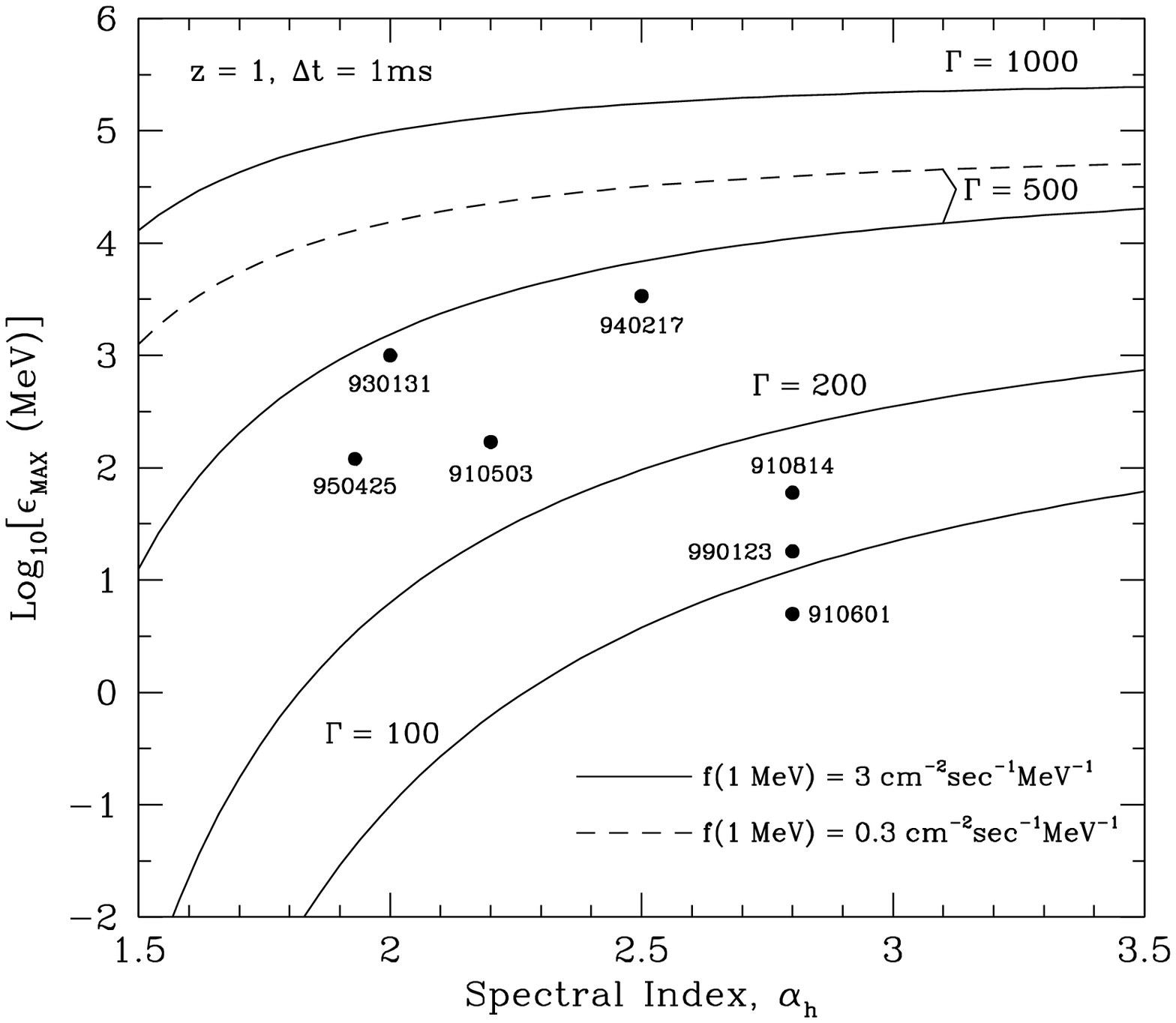}{3.8}{0.0}{-0.2}{
The phase space diagram for the observational parameters \teq{\Emax}
and \teq{\alpha_h}, consisting of contours of constant \teq{\Gamma}, as
labelled.  The contours correspond to solutions in
Eq.~(\ref{eq:Emax_GLAST}) of \teq{\taupp =1}.  Here \teq{z=1},
and a variability size of \teq{R=3\times 10^7\Gamma^2\,\hbox{cm}}
was adopted.  The solid and dashed (\teq{\Gamma=500} case only) 
contours correspond to the indicated choices of the 
flux \teq{f(\hbox{1 MeV)}} at 1 MeV.  The observed EGRET values of
\teq{\Emax} and \teq{\alpha_h} for the seven bursts in Table~1
are plotted as points; these bursts all have values of \teq{f(\hbox{1
MeV})} somewhat different from the chosen contour values. 
 \label{fig:Emax_phasespace}
}

The \teq{\Emax} curves are monotonically increasing functions of
\teq{\alpha_h} and \teq{\Gamma}, and are decreasing functions of
\teq{f(\hbox{1 MeV})}, as indicated by comparison of the solid and
dashed \teq{\Gamma =500} examples.  Since the optical depth is fixed at
unity at the turnover energy, and the flux is pinned at 1 MeV, thereby
providing a ``pivot point'' in the spectrum, increasing \teq{\alpha_h}
then lowers the number of interacting photons above pair threshold
(generally above 1 MeV in the observer's frame) so that \teq{\Emax}
correspondingly increases to compensate. The depiction of
\teq{(\alpha_h, \,\Emax )} positions for the EGRET bursts listed in
Table~1 clearly indicates that \teq{\Gamma\gtrsim 100} in their emission
regions if they are all at or near a redshift of \teq{z=1}. 
Observations of further extension of their power-laws by GLAST could
push these Lorentz factor bounds considerably higher: much of the phase
space below \teq{\Emax\sim 10^4}MeV and \teq{\alpha_h\lesssim 2.5}  is
readily accessible to GLAST, underlining its potential as a tool for
probing GRB bulk motion characteristics.  As the hard gamma-ray spectral
index increases to \teq{\alpha_h\gtrsim 2.8}, the total photon counts
collected by the LAT instrument will correspondingly be reduced,
lowering the maximum energies to which it can probe to below 1 GeV.

The results of Section~\ref{sec:evolution} highlight one and possibly
two groundbreaking probes of the burst environment by hard gamma-ray
experiments that explore the GLAST window.  First, the comparison of
Figs.~\ref{fig:GRB950425_spec} and~\ref{fig:GRB910503_spec} offers
strikingly different evolution of the turnover energy \teq{E_t} with
time (i.e., declining MeV-band flux) for constant \teq{\Gamma} and
evolving \teq{\Gamma} scenarios.  A sensitive hard gamma-ray detector
would easily be able to discern between these two signatures, using data
binned into several time windows, and cast light on the \teq{\Gamma}
evolution for a number of bright, energetic bursts.  Moreover, these
internal absorption signatures can be easily distinguished from external
absorption characteristics due to background fields, which generate
turnover energies independent of time or MeV-band flux.

This tracking of \teq{\Gamma} evolution is a global possibility that is
not confused by specific model nuances such as circumburst
non-uniformity and departures from temporal monotonicity for the flux. 
Should a burst exhibit \teq{\Gamma=}const. behavior, it can be concluded
that the expansion timescale is shorter than the dynamical one, so that
the burst duration \teq{T_{90}} in conjunction with Eq.~(\ref{eq:tdyn})
can be used to provide an approximate lower bound to the explosion
energy \teq{{\cal E}_0/\Omega_0} per unit solid angle of the expansion,
modulo the density of the environment surrounding the burst:
\begin{equation}
   \dover{{\cal E}_0}{\Omega_0}\;\gtrsim\; U_0\;\equiv\;
   \dover{1}{3}\, \Gamma_0^8\, \rho c^5\, T_{90}^3 \quad , \quad 
   \Gamma\; =\;\Gamma_0\; =\; \hbox{const}.
 \label{eq:erg_bound}
\end{equation}
On the other hand, should \teq{\Gamma} changes be observed, whether
increasing or declining with time, one could argue that the dynamical
timescale is smaller than the expansion age and that \teq{\Gamma <
\Gamma_0}.  Accordingly, an upper bound \teq{{\cal E}_0
/\Omega_0\lesssim U_0} on the collimated explosion energy would be
obtained, and this could be tightened to an approximate estimate \teq{
{\cal E}_0 /\Omega_0\sim \Gamma^8\rho c^5 T_{90}^3}, with considerable
uncertainty due to the spread in \teq{\Gamma} estimates derived from the
pair attenuation turnovers.  This calibration of \teq{{\cal
E}_0/\Omega_0} is analogous to inferences made in non-relativistic
supernova remnant expansions either in the coasting or Sedov phase.
Furthermore, if a radiative regime is deduced to be operating, then hard
gamma-ray observations can also constrain pertinent environmental
parameters for a chosen radiation mechanism, via bounds on the cooling
timescale.  Notwithstanding, it is noted that marked variations of
\teq{\alpha_h} with time can degrade this diagnostic possibility. Such a
property is evinced in spectroscopy analyses of some bursts (see, for
example, the joint BATSE/EGRET TASC study of Gonzalez et al. 2004),
though the average characteristics of time-resolved \teq{\alpha_h}
determinations by BATSE are similar to the time-integrated
\teq{\alpha_h} values (e.g. Kaneko et al. 2006).  GLAST will solidify
the understanding of temporal variations, or otherwise, in
\teq{\alpha_h} for many bursts.

An interesting, though somewhat less promising diagnostic is the
discrimination between adiabatic and radiative/non-adiabatic
evolutionary scenarios. While the evolutionary loci of \teq{E_t} for
these cases are distinct in Fig.~\ref{fig:GRB910503_spec}, they are only
modestly so, and can easily be blurred by breadth of the turnover, the
influence of non-uniformities in the burst environs, or complicated
variability in the spectral shape with time.  Furthermore, significant
changes in the value of the index \teq{\varrho} in
Eq.~(\ref{eq:eturn_locus_evolve}) are incurred in going from
longitudinal (\teq{\lambda =2}) to transverse variability (\teq{\lambda
=1}) situations. Hence it becomes harder to disentangle a variety of
effects given an observed, evolving, GeV-band spectrum.  Degeneracies
can be partially removed by exploring correlations with the MeV-band
peak evolution, in both flux and break energy.  For example,
Eqs.~(\ref{eq:eb_evolve}) and (\ref{eq:peakflux_evolve}) combine to give
a locus of breaks whose slope depends only on \teq{\eta_r}, and is
independent of \teq{\alpha_h}.  How effective such a correlated
diagnostic will be will depend on the sharpness of the respective
breaks.

GLAST's Large Area Telescope (LAT) offers potential for exploring these
signatures, being considerably more sensitive than EGRET; its principal
capability will be to provide lower bounds to \teq{\Gamma} via its
refinement of \teq{\Emax}, and perhaps also the first observations of
turnovers in the brightest subset of the burst population (e.g. see
Omodei et al. 2006, for simulated spectra with detectable turnovers).
Band, et al. (2004) and Cohen-Tanugi, et al. (2004) provide brief
outlines of the anticipated performance of GLAST for burst detections. 
Being ephemeral in nature, GRBs are vastly different from steady
sources, for which the LAT Performance Web Page {\tt
http://www-glast.slac.stanford.edu/software/IS/glast
\_lat\_performance.htm} offers putative integral sensitivities for the
\teq{5\sigma} detection of a high latitude source above a given energy,
resulting from a one-year, all-sky survey.   With the average 2
steradian field of view for the LAT, this corresponds to around
\teq{t_{\hbox{\sixrm 1yr}}=5\times 10^6}seconds observing a given
source, far above the typical duration of bursts.   For steady sources,
background is an issue (especially in the galactic plane: see Hunter et
al. 1997, for the presentation of EGRET background determinations), but
its influence on the observability for bursts virtually disappears since
they are of such short duration.  Gamma-ray burst detectability by the
LAT correlates to the total number of photons counted above its
threshold energy of 30 MeV.  Potential instrumental deadtime due to data
readout after triggers is small, about 5\%.  LAT Team simulations of
detector sensitivities (Omodei, et. al. 2006) indicate that only a
handful of bursts seen by GLAST each year will provide several hundred
photon counts in the LAT above the threshold of 30 MeV, and perhaps one
a year with of the order of a 1000 or so (and around one a year with 100
photons above 1 GeV).  Such cases could, in principal, yield of the
order of 30--100 photons above 300 MeV if the burst possesses a flat
\teq{E^{-2}} spectrum.  Observing turnovers around 100 MeV--1 GeV would
then be viable with such a database, and moreover, it may prove possible
to track spectral evolution at around 100 MeV--1 GeV by binning data in
two or three time intervals, with corresponding flux reductions by a
factor of 2--5 in the MeV band (or at higher energies).

Such is the optimal situation. Nature, of course, may not provide
turnovers at such low energies, so GLAST's potential for casting light
on the above bulk Lorentz factor diagnostics will only truly be known
after launch.  Yet, bright EGRET bursts typically have flat spectra, on
average around \teq{E^{-2}} (see Dingus, 1995, for a compilation
encompassing most of the bursts considered in this paper), which may or
may not be the result of flux selection effects at higher energies. 
These are ideal candidates for exploring spectral turnovers in the
super-GeV band, which were simulated by the LAT Collaboration in Omodei
et al. (2006). At best, only an elite handful of bursts over the GLAST
lifetime might be amenable to observing both turnovers in the sub-GeV
band, and partially tracking their evolution with time.  This number
will increase if the bulk Lorentz factors are actually slightly lower
than those used in the illustrations here. Accordingly, GLAST might be
capable of performing some of the GRB source diagnostics discussed in
the text. In addition, the diagnostics identified in this paper look to
the future, providing science motivations for next generation hard
gamma-ray detectors, both of the space-based variety and also
ground-based \v{C}erenkov arrays.  Since the atmospheric \v{C}erenkov
technique is intrinsically the more sensitive, due to its much greater
collection area, the current push to lower their  thresholds deep into
the GLAST band via increased telescope mirror size may inflluence the
direction gamma-ray burst observational programs targeting energies
above 1 GeV will take.  This advantage is offset by their generally
small fields of view (except for water tank \v{C}erenkov systems such as
MILAGRO), so that large fields of view like those afforded by the LAT
and the Gamma-Ray Burst Monitor (GBM) on GLAST are desirable.

\section{CONCLUSION}
 \label{sec:conclusion}

This paper outlines the interesting potential the GLAST mission and
future hard gamma-ray detectors can offer for probing the relativistic
nature of outflow in gamma-ray burst prompt emission regions. The
diagnostics highlighted here use a single, straightforward piece of
physics, namely that of electron-positron pair creation, and are only
dependent on model assumptions to second order.   Spectral attenuation
by this process, while not evident in EGRET data for a handful of
bright, hard bursts, should emerge from the GLAST database, unless the
bulk Lorentz factors in the burst emission zones are higher than
presently argued, i.e. greater than around \teq{\Gamma\sim 10^3}.  Then,
attenuation by interaction of source photons with those of the
intervening cosmic background field will produce attenuation turnovers
at above 30--50 GeV, whose energies are independent of the GRB flux at a
given time.  In such cases, bursts can be used to explore the evolution
of the background field with redshift by detectors with sufficient
sensitivity in this band.

For the equally interesting possibility of lower \teq{\Gamma}, pair
creation turnovers spawned by photons internal to the burst, will become
prominent in the GLAST energy window.  In such cases, the representative
calculations in Section~\ref{sec:evolution_effect} clearly illustrate
how the LAT instrument on GLAST might potentially discriminate between
evolutionary scenarios with constant \teq{\Gamma}, and those where
\teq{\Gamma} declines with time, should Nature be agreeable and offer
bright, hard bursts with turnovers below 300 MeV--1 GeV.  A more
sensitive, next-generation hard gamma-ray experiment might additionally
be capable of distinguishing between adiabatic and radiative
evolutionary scenarios: due to their different \teq{\Gamma (t)}
dependence, their turnover energy/flux loci trace somewhat different
tracks over time.  However, this probe will be more reliable for flat
spectrum bursts and require relatively sharp turnovers, or correlated
information from the MeV-band break evolution. While the GBM on GLAST,
in conjunction with the LAT, while accumulate a host of data on the
MeV-band break evolution, next-generation telescopes most likely will be
needed to differentiate the adiabatic and radiative turnover tracks.
Should distinction between these two keystone classes of blast wave
deceleration be possible, the pair creation probes offer constraints on
the explosion energy generated by the central engine.  Note that while
the results presented focus on simple broken power-law source spectra,
they are readily extended to apply to multiple components.  They
encompass a variety of possibilities, and provide a template for burst
pair production attenuation studies in the soon-to-be realized GLAST
era, and for future initiatives down the line.

\acknowledgments
I thank Alice Harding for many interactions over the years on pair
attenuation in bursts, for carefully reading the manuscript and
providing incisive suggestions for the clarification of certain points.
I thank Seth Digel for extensive discussions on GLAST/LAT sensitivity
expectations, Jay Norris and David Band for insights into GLAST/LAT
capabilities for gamma-ray burst detections, and Brenda Dingus for
reading through the paper and providing updates on atmospheric
\v{C}erenkov telescope observations of GRBs.  I also thank the referee,
Peter M\'esz\'aros for recommendations helpful in the polishing of the
manuscript. This research was supported in part by the National Science
Foundation under Grant Nos.~AST00-98705 and PHY99-07949, and I
acknowledge the hospitality of the Kavli Institute for Theoretical
Physics, University of California, Santa Barbara, where part of the work
for this paper was performed.

\appendix
\section{Integration of the Angular Contributions to \teq{\taupp (\erg)} for Strong Beaming}
 \label{sec:appendix}
Here the analytic reduction of the triple integration in
Eq.~(\ref{eq:calFapprox}) is outlined, together with relevant properties
of the emergent hypergeometric function. It is expedient to scale the
integration variables via \teq{x_{\omega} = w x_{\erg}} and \teq{\kappa
= x_{\erg}\rho}, so that \teq{x_{\erg}}, \teq{w} and \teq{\rho}
constitute the new set of integration variables. Then, one performs the
\teq{w} integration first, which is easily tractable, and generates a
simple evaluation in terms of inverse trigonometric functions using
identity 2.261 of Gradshteyn \& Ryzhik (1980).  The \teq{x_{\erg}}
integration is then trivial, absorbing the constraints imposed by the
step function \teq{\Theta [\, \kappa\eta\theta_m/(2\chi)\, ]}.  The
result can then be expressed (for \teq{0\leq\Psi\leq 1}) as
\begin{equation}
   {\cal F}_{\alpha} \; \approx\; \theta_m^{2\alpha}\; 
        {\cal G}_{\alpha}(\Psi )\quad ,\quad 
        {\cal G}_{\alpha}(\Psi )\; =\; \dover{16}{\pi (\alpha + 2)}\;
                \Biggl\{ \int_0^{\Psi} dy\; y^{1+2\alpha}\,\arccos y
               + \Psi^{2(\alpha + 2)} \int_{\Psi}^1 \dover{dy}{y^3}\,\arccos y\;\Biggr\}\quad ,
 \label{eq:Falp_eval1}
\end{equation}
with now the one parameter
\begin{equation}
   \Psi \; =\; \dover{\chi}{\eta\theta_m}
 \label{eq:Omega_def_app}
\end{equation}
defining the general character of \teq{{\cal F}_{\alpha}}.  From this
form, the substitution \teq{y=\sin\beta} and the invocation of
identities 3.621 and 8.335.1 of Gradshteyn \& Ryzhik (1980) lead to the
result in Eq.~(\ref{eq:calAdef}) when \teq{\Psi =1}. On the other hand,
the \teq{\Psi\ll 1} limit is readily discerned, thereby reproducing
Eq.~(\ref{eq:calF_largeeta}).  These two cases correspond to the results
\begin{equation}
    {\cal G}_{\alpha}(1)\; =\; {\cal A}(\alpha )\;\; ,\quad \hbox{and} \quad
    {\cal G}_{\alpha}(\Psi)\; \sim\; \dover{4}{\alpha + 1}
                        \; \Psi^{2(1+\alpha )}\quad ,\quad \Psi\;\ll\; 1\quad .
 \label{eq:calG_asymp}
\end{equation}
For \teq{\Psi > 1}, corresponding to cases where the angular
integrations are completely separated form the threshold constraint,
\teq{{\cal G}_{\alpha}(\Psi )={\cal G}_{\alpha}(1)={\cal A}(\alpha )}.
For general \teq{0\leq\Psi < 1}, the integrals for \teq{{\cal
G}_{\alpha}(\Psi)} can be manipulated by the appropriate integration by
parts, and then the second integral term in Eq.~(\ref{eq:Falp_eval1}) is
routinely handled.   The first integral is somewhat more complicated,
and can be developed by using the change of variables \teq{y=\Psi\cos x}
together with the aid of the result 3.671 of Gradshteyn \& Ryzhik
(1980).  This renders the first integral expressible in terms of the
ordinary hypergeometric function \teq{F\equiv {}_2F_1}, so that
\begin{equation}
   {\cal G}_{\alpha}(\Psi)\; =\; \dover{8}{\pi}\, \Psi^{2(1+\alpha )}
        \Biggl\{ \dover{\arccos\Psi}{\alpha +1}
                      - \dover{\Psi\sqrt{1-\Psi^2}}{\alpha +2}
                      +\dover{\Psi\, F(1/2,\, \alpha+ 3/2;\, \alpha + 5/2;\, \Psi^2)}{
                                     (\alpha +1)\, (2\alpha +3)\, (\alpha +2)} \Biggr\}\quad , \quad
                                     0\;\leq\;\Psi\;\leq\; 1\quad .
 \label{eq:calG_final}
\end{equation}
This hypergeometric function reduces to combinations of inverse
trigonometric and algebraic functions when \teq{\alpha} assumes integer
values, and just algebraic functions when \teq{\alpha} takes on half
integral values. For this range of \teq{\Psi}, the hypergeometric
function can be numerically evaluated using the series expansion 9.100
of Gradshteyn \& Ryzhik (1980), yielding
\begin{equation}
    F\Bigl[\, \dover{1}{2},\, \alpha + \dover{3}{2};\, \alpha + \dover{5}{2};\, z\,\Bigr]
             \; =\; \sum_{n=0}^{\infty} a_nz^n\quad ,\quad
    a_n\; =\; \dover{2\alpha+3}{2\alpha + 2n+3}\; 
                    \prod_{k=1}^n \biggl( 1 - \dover{1}{2k} \biggr)\quad .
 \label{eq:hypergeom_series}
\end{equation}
For \teq{0\leq z < 1}, this series converges rapidly, faster than a
geometric series. In the boundary case of \teq{z=1}, convergence is 
only moderately rapid.  The asymptotic behavior of the product in
Eq.~(\ref{eq:hypergeom_series}) as \teq{n\to\infty} can be established
by taking its logarithm and then using the series identity 44.9.1 of
Hansen (1975):
\begin{equation}
   \sum_{k=1}^{\infty} \biggl\lbrack \, \dover{x}{k}
                                - \log_e \Bigl(1 + \dover{x}{k} \Bigr)\,\biggr\rbrack
          \; =\; \gamma_{\hbox{\fiverm E}} x + \log_e\Gamma(x+1)\quad ,
 \label{eq:Hansen_44.9.1}
\end{equation}
where \teq{\Gamma (z)} is the Gamma function, and
\teq{\gamma_{\hbox{\fiverm E}} =-\psi (1)\approx 0.57721} is the
Euler-Mascheroni constant, with \teq{\psi (z) = d/dz \{\log_e\Gamma(z)\} }.  
By truncating the resultant infinite series and then invoking
results 8.365.3 and 8.365.5 of Gradshteyn \& Ryzhik (1980), the
\teq{n\to\infty} limit can eventually be taken, thereby deriving the
result
\begin{equation}
   \lim_{n\to\infty} \sqrt{n}\; \prod_{k=1}^n \biggl( 1 - \dover{1}{2k} \biggr)
             \; =\; \dover{1}{\sqrt{\pi}}\quad .
 \label{eq:product_limit}
\end{equation}
Hence, the series expansion in Eq.~(\ref{eq:hypergeom_series}) converges
as \teq{n^{-3/2}} for large \teq{n}.  Such a rate of convergence is not
compelling, so that for values \teq{z\approx 1} it is more expedient to
perform the transformation indicated in 9.131.2 of Gradshteyn \& Ryzhik
(1980), and then use the resultant identity
\begin{equation}
   F\Bigl[\, \dover{1}{2},\, \alpha + \dover{3}{2};\, \alpha + \dover{5}{2};\, z\,\Bigr]
   \; =\; \sqrt{\pi}\; \dover{\Gamma (\alpha + 5/2)}{\Gamma (\alpha +2)}\;
            F\Bigl[\, \dover{1}{2},\, \alpha + \dover{3}{2};\, \dover{1}{2};\, 1-z\,\Bigr]
   -  (2\alpha + 3)\; \sqrt{1-z}\;
            F\Bigl[\, 1,\, \alpha + 2\,;\, \dover{3}{2};\, 1-z\,\Bigr]   
 \label{eq:hypergeom_ident}
\end{equation}
to evaluate the left hand side for domains of \teq{(1-z)\ll 1}.  The
rate of convergence of the two hypergeometric functions on the right
hand side of Eq.~(\ref{eq:hypergeom_ident}) is then at least as rapid as
a geometric series.  This transformation also expeditiously establishes
the correspondence \teq{{\cal G}_{\alpha}(\Psi )\to {\cal A}(\alpha )}
as \teq{\Psi\to 1^-}, using the doubling formula (8.335.1 of Gradshteyn
\& Ryzhik 1980) for the Gamma function.  For numerical purposes, it is
obviously convenient to use Eq.~(\ref{eq:hypergeom_series}) for
\teq{0\leq z\leq 1/2} cases, and the power series expansion about
\teq{z=1} of the right hand side of Eq.~(\ref{eq:hypergeom_ident}) when
\teq{1/2 < z\leq 1}.


To conclude, two integrals that are required for the determination of
the asymptotic approximation for \teq{{\cal T}(\theta_m ,\, \eta )} when
\teq{\eta\theta_m\gg 1} are evaluated.  Using the integral form for
\teq{{\cal G}_{\alpha}(\Psi )} in Eq.~(\ref{eq:Falp_eval1}), it can be
quickly demonstrated that
\begin{equation}
   \int_0^1 \dover{d\Psi}{\Psi^{1+2\alpha}}\; {\cal G}_{\alpha}(\Psi )
       \; =\; \dover{1-{\cal A}(\alpha )}{2\alpha} \quad ,\quad
    \int_0^1 d\Psi\; \dover{\log_e\Psi}{\Psi^{1+2\alpha}}\; {\cal G}_{\alpha}(\Psi )
       \; =\; \dover{1-{\cal A}(\alpha )}{4\alpha^2} - \dover{4\log_e2+1}{8\alpha}
\label{eq:int_calG_idents}
\end{equation}
by appropriate integration by parts to generate the derivative 
\teq{d{\cal G}_{\alpha}/d\Psi} in each case.


\begin{references}

\reference{}
   Aharonian, F., et al. 2006, \nat,\vol{440}{1018}
\reference{} 
   Albert, J., et al. 2006a, \apjl,\vol{641}{L9}
\reference{} 
   Albert, J., et al. 2006b, \apjl,\vol{642}{L119}
\reference{} 
   Atkins, R., et al. 2000, \apjl,\vol{533}{L119}
\reference{} 
   Atkins, R., et al. 2003, \apj,\vol{583}{824}
\reference{} 
   Atkins, R., et al. 2005, \apj,\vol{630}{996}
\reference{} 
   Band, D.~L., et al. 1993, \apj,\vol{413}{281}
\reference{} 
   Band, D.~L., et al. 2004, in {\it Gamma-Ray Bursts: 30 years of Discovery}, eds.
   Fenimore, E.~E. \& Galassi, M. (AIP Conf. Proc. 727, New York) p.~692.
\reference{} 
   Baring, M.~G. 1993, \apj,\vol{418}{391}
\reference{} 
   Baring, M.~G. 1994, \apjs,\vol{90}{899}
\reference{}
   Baring, M.~G. 2001, in {\it Gamma-Ray Astrophysics},
    eds. S. Ritz, N. Gehrels \& C. Schrader 
    (AIP Conf. Proc. 587, New York), p.~153.
\reference{} 
   Baring, M.~G. \& Braby, M.~L. 2004, \apj,\vol{613}{460}
\reference{}
   Baring, M.~G. and Harding, A.~K. 1996, in {\it Gamma-Ray Bursts}, eds.  
   Kouveliotou, C., Briggs, M.~S., and Fishman, G.~J. (AIP Conf. Proc. 384, 
   New York) p.~724.
\reference{} 
   Baring, M.~G. \& Harding, A.~K. 1997a, \apjl,\vol{481}{L85} (BH97a)
\reference{} 
   Baring, M.~G. \& Harding, A.~K. 1997b, \apj,\vol{491}{663}
\reference{}
   Bednarz, J. \& Ostrowski, M. 1998, \prl,\vol{80}{3911}
\reference{}
   Berger,ÊE., et al. 2006, \apj,\vol{642}{979}
\reference{}
   Blandford, R.~D. \& McKee, C.~F., 1976, Phys. Fluids, \vol{19}{1130}
\reference{} 
   Briggs, M.~S., et al. 1999, \apj,\vol{524}{82}
\reference{} 
   Catelli, J.~R., Dingus, B.~L. \& Schneid, E.~J. 1996, in {\it Gamma-Ray Bursts}, eds.  
   Kouveliotou, C., Briggs, M.~F., and Fishman, G.~J. (AIP Conf. Proc. 384, 
   New York) p.~158.
\reference{}
   Cohen-Tanugi, J., et al. 2004, in {it Third Rome Workshop on Gamma-Ray
   Bursts in the Afterglow Era}, eds. Feroci, M., Frontera, F., Masetti, N. \& Piro, L.
   (ASP Conf. Ser. 312) p.~520.
\reference{}
   Connaughton, V., et al. 1997, \apj,\vol{479}{859}
\reference{}
   Crider, A. \& Liang, E.~P. 1999, \aaps,\vol{138}{405}
\reference{}
   Cummings, J., et al. 2005, GCN 3910.
\reference{}
   Dermer, C.~D., Chiang, J. \& B\"ottcher, M. 1999, \apj,\vol{513}{656}
\reference{}
   Dermer, C.~D., Chiang, J. \& Mitman, K.~E. 2000, \apj,\vol{537}{785}
\reference{} 
   Dingus, B.~L. 1995, \apss,\vol{231}{187}
\reference{}
   Epstein, R.~I. 1973, \apj,\vol{183}{593}
\reference{}  
   Epstein, R.~I. 1985, \apj,\vol{297}{555}
\reference{}  
   Fenimore, E.~E., Epstein, R.~I. \& Ho, C.: 1992 in {\it Gamma-Ray Bursts},  
   eds. Paciesas, W. S. \& Fishman, G. J., (AIP Conf. Proc. 265, New York) p.~158.
\reference{}  
   Ford, L.~A., et al. 1995, \apj,\vol{439}{307}
\reference{}
   Gonz\'alez, M.~M, Dingus, B.~L., Kaneko, Y., Preece, R.~D.  \& Briggs, M.~S.
   2004,  in {\it Gamma-Ray Bursts: 30 Years of Discovery}, 
   eds. Fenimore, E.~E. \& Galassi, M., (AIP Conf. Proc. 727, New York) p.~236.
\reference{}  
   Gould, R.~J. \& Schreder, G.~P. 1967, \pr,\vol{155}{1404}
\reference{}
   Gradshteyn, I.~S. \& Ryzhik, I.~M. 1980, {\it Table of Integrals, Series
   and Products}, (Academic Press, New York).
\reference{}
   Haislip, J., et al. 2006, \nat,\vol{440}{181} 
\reference{}
   Hanlon, L.~O., et al. 1994, \aap,\vol{285}{161}
\reference{}
   Hansen, E.~R. 1975, {\it A Table of Series and Products},
   (Prentice-Hall, Englewood Cliffs) 
\reference{}
   Hjorth, J., et al. 1999, GCN Circular 219.
\reference{}
   Hunter, S.~D., et al. 1997, \apj,\vol{481}{205}
\reference{}
   Hurley, K., et al. 1994, \nat,\vol{372}{652}
\reference{}
   Jakobsson, P. et al. 2006, \aap,\vol{447}{897}
\reference{}  
   Jarvis, A., et al. 2005, Proc. 29th Int. Cosmic Ray Conf. (Pune), \vol{4}{455}
\reference{}  
   Kaneko, Y., et al. 2006, in {\it Gamma Ray Bursts in the Swift Era},
   eds. Holt, S.~S., Gehrels, N. \& Nousek, J. (AIP Conf. Proc., New York), p.~133.
\reference{}
   Katz, J.~I. 1994, \apjl,\vol{432}{L27}
\reference{}
   Kawai, N., et al. 2005, GCN 3937.
\reference{}
   Kirk, J.~G., Guthman, A.~W., Gallant, Y.~A., Achterberg, A. 2000,
   \apj,\vol{542}{235}
\reference{}
   Kneiske, T.~M., Mannheim, K. \& Hartmann, D. 2002, \aap,\vol{386}{1}
\reference{}
   Kouveliotou, C., et al. 1994, \apjl,\vol{422}{L59}
\reference{}  
   Krolik, J.~H. \& Pier, E.~A. 1991, \apj,\vol{373}{277}
\reference{}
   Kwok, P.~W., et al. 1993, in Compton Gamma-Ray Observatory, 
   eds. Friedlander, M., Gehrels, N., and Macomb, D. (AIP Conf. Proc. 280,
   New York) p.~855.
\reference{}
   Lithwick, Y. \& Sari, R.  2001, \apj,\vol{555}{540}
\reference{}
   Longair, M.~S. 1998, {\it Galaxy Formation},
   (Springer-Verlag, Berlin)
\reference{}
   MacMinn, D. \& Primack, J. 1996, \ssr,\vol{75}{413}
\reference{}  
   Mannheim, K., Hartmann, D. \& Funk, B. 1996, \apj,\vol{467}{532}
\reference{}  
   Meegan, C., et al. 1996, \apjs,\vol{106}{65}   
\reference{}  
   M\'esz\'aros, P. \& Rees, M.~J. 1994, \mnras,\vol{269}{L41}
\reference{}  
   M\'esz\'aros, P., Rees, M.~J. \& Wijers, R.~A.~M.~J. 1998, \apj,\vol{499}{301}
\reference{}  
   M\'esz\'aros, P., Rees, M.~J. \& Papathanassiou, H. 1994, \apj,\vol{432}{181}
\reference{}  
   Omodei, N., et al. 2006, to appear in {\it Gamma Ray Bursts in the Swift Era},
   eds. Holt, S.~S., Gehrels, N. \& Nousek, J. (AIP Conf. Proc., New York), p.~642.
\reference{}  
   Perlmutter, S., et al. 1997, \apj,\vol{483}{565}
\reference{}
   Preece, R. D., et al. 1995, \apss,\vol{231}{149} 
\reference{} 
   Preece, R.~D., et al. 1998, \apjl,\vol{506}{L23}
\reference{}
   Preece, R. D., et al. 2000, \apjs,\vol{126}{19} 
\reference{}
   Primack, J., Somerville, R.~S., Bullock, J.~S. \& Devriendt, J.~E.~G.
   2001, in {\it High Energy Gamma-Ray Astronomy}, eds. F. Aharonian \&
   H. V\"olk, (AIP Conf. Proc. 558, New York), p.~463.  
\reference{} 
   Rees, M.~J. \& M\'esz\'aros, P. 1992, \mnras,\vol{258}{41P}
\reference{}  
   Riess, A.~G., et al. 1998, \aj,\vol{116}{1009}
\reference{}
   Sari, R. \& Esin, A. 2001, \apj,\vol{548}{787}
\reference{}  
   Schaefer, B.~E., et al. 1992, \apjl,\vol{393}{L51}
\reference{}  
   Schmidt, W.~K.~H. 1978, \nat,\vol{271}{525}
\reference{}  
   Schneid, E.~J. 1992, \aapl,\vol{255}{L13}
\reference{}
   Sommer, M., et al. 1994, \apjl,\vol{422}{L63}
\reference{}
   Spergel, D.~N., et al. 2003, \apjs,\vol{148}{175}
\reference{}
   Stecker, F.~W. 2001, in Proc. IAU Symposium 204, eds. M. Harwit \& 
   M. G. Hauser, p.~135. 
\reference{}
   Stecker, F.~W. \& De Jager, O.~C. 1996, \ssr,\vol{75}{401}
\reference{}
   Stecker, F.~W., De Jager, O.~C. \& Salamon, M.~H. 1996, \apjl,\vol{390}{L49}
\reference{}
   Stepney, S. \& Guilbert, P.~W. 1983, \mnras,\vol{204}{1269}
\reference{}
   Svensson, R. 1987, \mnras ,\vol{227}{403}
\reference{}
   Woods, E. \& Loeb, A. 1995, \apj,\vol{453}{583}

\end{references}
\end{document}